\documentclass[10pt,journal,compsoc]{IEEEtran}
\ifCLASSOPTIONcompsoc
  \usepackage[nocompress]{cite}
\else
  \usepackage{cite}
\fi
\ifCLASSINFOpdf
\else
\fi
\usepackage{hyperref}
\usepackage{microtype}
\usepackage{amsmath}
\usepackage{amsthm}
\usepackage{multirow}
\usepackage{graphicx}
\usepackage{subfigure}
\usepackage{booktabs} % for professional tables
\graphicspath{{./figures/}} % include the picture folder in here

\newtheorem{theorem}{Theorem}
\newtheorem{definition}{Definition}
\usepackage{xcolor}
\newcommand{\dv}[1]{{\color{black}{#1}}}

\begin{document}
%
% paper title
% Titles are generally capitalized except for words such as a, an, and, as,
% at, but, by, for, in, nor, of, on, or, the, to and up, which are usually
% not capitalized unless they are the first or last word of the title.
% Linebreaks \\ can be used within to get better formatting as desired.
% Do not put math or special symbols in the title.
\title{Privacy-preserving Deep Learning based Record Linkage}
%
%
% author names and IEEE memberships
% note positions of commas and nonbreaking spaces ( ~ ) LaTeX will not break
% a structure at a ~ so this keeps an author's name from being broken across
% two lines.
% use \thanks{} to gain access to the first footnote area
% a separate \thanks must be used for each paragraph as LaTeX2e's \thanks
% was not built to handle multiple paragraphs
%
%
%\IEEEcompsocitemizethanks is a special \thanks that produces the bulleted
% lists the Computer Society journals use for "first footnote" author
% affiliations. Use \IEEEcompsocthanksitem which works much like \item
% for each affiliation group. When not in compsoc mode,
% \IEEEcompsocitemizethanks becomes like \thanks and
% \IEEEcompsocthanksitem becomes a line break with idention. This
% facilitates dual compilation, although admittedly the differences in the
% desired content of \author between the different types of papers makes a
% one-size-fits-all approach a daunting prospect. For instance, compsoc 
% journal papers have the author affiliations above the "Manuscript
% received ..."  text while in non-compsoc journals this is reversed. Sigh.

\author{Thilina~Ranbaduge,
        Dinusha~Vatsalan,%~\IEEEmembership{Fellow,~OSA,}
        ~and~Ming~Ding,~\IEEEmembership{~Senior Member,~IEEE}% <-this % stops a space
\IEEEcompsocitemizethanks{
\IEEEcompsocthanksitem T. Ranbaduge and M. Ding are with 
the Commonwealth Science and Industrial Research Organisation (CSIRO), Australia.
\protect\\
% note need leading \protect in front of \\ to get a newline within \thanks as
% \\ is fragile and will error, could use \hfil\break instead.
E-mail: thilina.ranbaduge@data61.csiro.au, ming.ding@data61.csiro.au.
\IEEEcompsocthanksitem D. Vatsalan is with Macquarie University, NSW, Australia.
\protect\\
E-mail: dinusha.vatsalan@mq.edu.au.
}% <-this % stops an unwanted space
\thanks{Manuscript received April 19, 2005; revised August 26, 2015.}}

% note the % following the last \IEEEmembership and also \thanks - 
% these prevent an unwanted space from occurring between the last author name
% and the end of the author line. i.e., if you had this:
% 
% \author{....lastname \thanks{...} \thanks{...} }
%                     ^------------^------------^----Do not want these spaces!
%
% a space would be appended to the last name and could cause every name on that
% line to be shifted left slightly. This is one of those "LaTeX things". For
% instance, "\textbf{A} \textbf{B}" will typeset as "A B" not "AB". To get
% "AB" then you have to do: "\textbf{A}\textbf{B}"
% \thanks is no different in this regard, so shield the last } of each \thanks
% that ends a line with a % and do not let a space in before the next \thanks.
% Spaces after \IEEEmembership other than the last one are OK (and needed) as
% you are supposed to have spaces between the names. For what it is worth,
% this is a minor point as most people would not even notice if the said evil
% space somehow managed to creep in.

% The paper headers
\markboth{IEEE Transactions Journals}%
{Ranbaduge \MakeLowercase{\textit{et al.}}: Bare Demo of IEEEtran.cls for Computer Society Journals}
% The only time the second header will appear is for the odd numbered pages
% after the title page when using the twoside option.
% 
% *** Note that you probably will NOT want to include the author's ***
% *** name in the headers of peer review papers.                   ***
% You can use \ifCLASSOPTIONpeerreview for conditional compilation here if
% you desire.

% The publisher's ID mark at the bottom of the page is less important with
% Computer Society journal papers as those publications place the marks
% outside of the main text columns and, therefore, unlike regular IEEE
% journals, the available text space is not reduced by their presence.
% If you want to put a publisher's ID mark on the page you can do it like
% this:
%\IEEEpubid{0000--0000/00\$00.00~\copyright~2015 IEEE}
% or like this to get the Computer Society new two part style.
%\IEEEpubid{\makebox[\columnwidth]{\hfill 0000--0000/00/\$00.00~\copyright~2015 IEEE}%
%\hspace{\columnsep}\makebox[\columnwidth]{Published by the IEEE Computer Society\hfill}}
% Remember, if you use this you must call \IEEEpubidadjcol in the second
% column for its text to clear the IEEEpubid mark (Computer Society jorunal
% papers don't need this extra clearance.)

% use for special paper notices
%\IEEEspecialpapernotice{(Invited Paper)}

% for Computer Society papers, we must declare the abstract and index terms
% PRIOR to the title within the \IEEEtitleabstractindextext IEEEtran
% command as these need to go into the title area created by \maketitle.
% As a general rule, do not put math, special symbols or citations
% in the abstract or keywords.
\IEEEtitleabstractindextext{%
\begin{abstract}
\emph{Deep learning}-based linkage of records across different databases is becoming increasingly useful in data integration and mining applications to discover new 
insights from multiple sources of data. 
However, 
due to privacy and confidentiality concerns, 
organisations often are not willing or allowed to share their sensitive data with any external parties, 
thus making it challenging to build/train deep 
learning models for record linkage across different organizations' databases. 
To overcome this limitation, 
we propose the first deep learning-based multi-party 
privacy-preserving record linkage (PPRL) protocol that can be used to link sensitive databases held by multiple different organisations. 
In our approach, 
each database owner first trains a local deep learning model, 
which is then uploaded to a secure environment and securely aggregated to create a global model. 
The global model is then used by a linkage unit to distinguish unlabelled record pairs as matches and 
non-matches. 
We utilise differential privacy to achieve provable privacy protection against re-identification attacks. 
We evaluate the linkage quality and scalability 
of our approach using several large real-world databases, showing that it can achieve high linkage quality while providing sufficient privacy protection against existing
attacks.
\end{abstract}

% Note that keywords are not normally used for peerreview papers.
\begin{IEEEkeywords}
Differential privacy, deep neural networks, data integration, Bloom filter encoding.
\end{IEEEkeywords}}

% make the title area
\maketitle

% To allow for easy dual compilation without having to reenter the
% abstract/keywords data, the \IEEEtitleabstractindextext text will
% not be used in maketitle, but will appear (i.e., to be "transported")
% here as \IEEEdisplaynontitleabstractindextext when the compsoc 
% or transmag modes are not selected <OR> if conference mode is selected 
% - because all conference papers position the abstract like regular
% papers do.
\IEEEdisplaynontitleabstractindextext
% \IEEEdisplaynontitleabstractindextext has no effect when using
% compsoc or transmag under a non-conference mode.

% For peer review papers, you can put extra information on the cover
% page as needed:
% \ifCLASSOPTIONpeerreview
% \begin{center} \bfseries EDICS Category: 3-BBND \end{center}
% \fi
%
% For peerreview papers, this IEEEtran command inserts a page break and
% creates the second title. It will be ignored for other modes.
\IEEEpeerreviewmaketitle

\IEEEraisesectionheading{\section{Introduction}
\label{sec:introduction}}

Organisations in many business domains increasingly produce large databases with
millions of records, which might contain detailed and sensitive information about
people, such as customers, patients, taxpayers, or travellers. Often such databases
need to be shared and integrated to facilitate advanced analytics and processing. 
However, due to privacy and confidentiality concerns, organisations are not allowed
or not willing to share their databases in plain text 
%\dv{in their original form} 
for data linkage purposes~\cite{Chr2020}. 

In this light, \emph{privacy-preserving record linkage} (PPRL) aims to develop
techniques that facilitate the linking of databases without the need of any 
sensitive data to be shared among the involved organisations~\cite{Chr2020,Vat13}.
%privacy-preserving record linkage (PPRL) becomes useful because it identifies
%and matches records that refer to the same entity across sensitive databases held by 
%different organisations \dv{without compromising sensitive or private information about 
%individuals in the record linkage process}~\cite{Chr2020, Vat13}.
This process is often challenging because no unique entity identifiers, 
such as social security numbers, are available in the databases to be linked. 
Therefore, quasi-identifying attributes such as names and addresses, 
are used to identify records that are similar and likely belong to the same entity~\cite{Chr12}. Such quasi-identifiers are however often not allowed to be 
shared between organisations due to privacy and confidentiality concerns. 

A popular approach to link sensitive data in a privacy-preserving way is to encode quasi-identifying values, such as using Bloom filters, that allows 
\dv{fuzzy matching by calculating approximate similarities on the encoded values 
in order to identify matches and non-matches}~\cite{Chr2020, dur13, Vat13}. 
Traditional PPRL applications commonly use a na\"ive threshold-based classifier
to classify encoded record pairs as matches if their corresponding approximate
similarity scores are above an user defined similarity threshold~\cite{dur13, Vat13}. 
Recent record linkage literature has shown that supervised classifiers, 
such as deep learning techniques, can yield significantly high accuracy of 
linkage, especially in big data applications, where the data is prone to 
errors and variations~\cite{Ebr18,Koo18,xue2020}.

%
%\dv{Classifying records as matches and non-matches based on the approximate similarity scores 
%has traditionally been performed by using a simple threshold-based classifiers~\cite{Vat13}. 
%These threshold-based classifiers classify the record pairs as matches if the calculated 
%similarity score between their quasi-identifiers is above a certain similarity threshold. 
%However, as studied in the record linkage (also known as entity resolution) literature, 
%more advanced classifiers such as using machine learning or deep learning techniques to
%classify record pairs can yield significantly high accuracy of linkage, especially in Big 
%data applications, where the data is prone to errors and variations~\cite{Di19,Ebr17,Koo18,xue2020}}.
%
\dv{%In contrast to traditional
%record linkage, PPRL processes are often more challenging due to the privacy-preserving requirements. 
However, 
deep learning brings in several challenges for PPRL.}
%DV: need to discuss these challenges!
% I think in general we can say lack of training data is an issue for 
% supervised record linkage and in PPRL sharing of such training data
% is not always possible due to privacy requirements. 
%So I think the first few sentences can be rephrase as, 
%Supervised linkage approaches have recently gained much attention due to 
% the high quality results that can be obtained when sufficient training data
%are available. Such approaches, however, often require extensive human
%efforts to label record pairs as matches and non-matches. Such manual
%labelling is costly and time consuming, making it infeasible in many
%real-world applications.
%
First, supervised techniques, such as deep learning, require large amounts of 
training data to train a linkage model, which is challenging due to the lack 
of ground truth links. Further, generating training data with manual labelling
is costly and time consuming, leading to limited available training data. 
Second, training a global linkage model across multiple databases 
(held by different parties) requires exchanging the training data of individual
parties among each other or with a trusted linkage unit, which raises additional
security and privacy concerns.

%not available in the first place due
%to lack of data, mainly caused by privacy and confidentiality concerns as discussed 
%before~\footnote{Also, generating training data with manual labelling is costly and 
%time consuming, making it infeasible in many real-world applications}. Second, 
%training a global linkage model for multiple databases (held by different parties) 
%requires exchanging the training data of individual parties among each other or with 
%a trusted linkage unit, which raises additional security and privacy concerns. }

%\md{[Ming]: I suppose further explanation is needed for the following sentence.}
% Dinusha address this. 

Moreover, the use of a distributed learning setting, such as federated learning, 
in the PPRL context might not overcome the problem of exchanging sensitive data
entirely %issue of lack of training data in deep learning entirely 
due to the privacy risk involved in the linkage process. This is because the 
database owners need to reveal their unlabelled records to calculate 
record pair similarities with other database owners' unlabelled records.
%This is
%because individual encoded records are required to compute the similarity scores
%of record pairs from different databases in the classification process. 
Thus, privacy techniques need to be applied to individual records instead of 
the local models as performed in federated learning. To the best of our knowledge,
deep learning has not been studied for PPRL so far, while its advantages have 
been effectively utilized in non-PPRL applications. 

%\dv{Moreover, deep learning-based linkage cannot be done in a federated learning 
%environment that can reduce the privacy risk of revealing records of individual parties. 
%since individual records are required (in the encoded and/or encrypted form) to 
%compute the similarity scores of record pairs from different parties and to 
%classify as matches or not based on the similarity scores, the linkage operation 
%cannot be done using a federated learning model, where the parties do not have 
%to reveal their data, but only reveal locally trained models to compute a global 
%model. This means, privacy-preservation needs to be considered in the individual 
%record level instead of model-level that is trained on collective records. 
%To the best of our knowledge, deep learning-based classifiers have not
%been studied for PPRL so far.}

\textbf{Contribution:} \dv{In this work, 
we aim to investigate the application of deep learning in the PPRL context.} 
%We propose a novel protocol for 
%PPRL using deep learning classifiers and investigate the applicability in a 
%privacy-preserving context.}
%
%because quasi-identifiers in
%different databases are not allowed to be shared between organisations due to 
%privacy and confidentiality concerns. Thus, we aim to apply deep learning techniques
%to the encoded data to classify record pairs into matches and non-matches. 
%
\dv{The main contributions of this work are:}
%\dv{
%\begin{enumerate}
%    \item We propose a novel multiparty linkage protocol for deep-learning based PPRL
%    \item We theoretically study the privacy guarantees for such linkage models 
%    with Differential privacy
%    \item We empirically evaluate the linkage quality, efficiency, and privacy 
%    guarantees using nine different data sets. We evaluate
%    the privacy guarantees against a recently proposed attack method~\cite{Chr17}.
%    \item We compare our protocol with three  baseline methods, which are 1) PPRL with 
%    Bloom filter encoding that is not differentially private and with no deep learning, 
%    2) PPRL with Bloom filter encoding that is not differentially private, but with deep
%    learning, and 3) PPRL with differentially private Bloom filter encoding, but no 
%    deep learning, and compare with our proposed method, i.e. PPRL with differentially 
%    private Bloom filter encoding and with deep learning.
%\end{enumerate}
%}
(1) we propose a novel and first PPRL protocol that can be used to link multiple databases with limited training data using a deep learning model, 
(2) we utilise differentially private Bloom filters in the linkage process to provide privacy guarantees for entities associated with the records in the databases, 
(3) we theoretically prove the privacy guarantees provided by our approach, 
and (4) empirically evaluate the linkage quality, efficiency, and privacy guarantees using nine data sets from different domains, 
which shows that our approach can defend against a recently proposed PPRL attack method~\cite{Chr17,chr18a}, 
while achieving a high linkage quality compared to several baselines.

\dv{
\textbf{Outline:} The rest of the paper is organised as follows: 
We review related work in the following section and provide preliminaries in Section~\ref{sec:background}. In Section~\ref{sec:method}, we describe our
proposed protocol for PPRL based on deep learning classifiers and analyse
the privacy guarantees of our protocol in 
Section~\ref{sec:analysis}. We present and discuss the results of our 
experimental study in Section~\ref{sec:experiments}. Finally, we conclude 
and point out directions of future research in Section~\ref{sec:conclusion}.
}

%TR: Outline can be removed if we need to shorten the paper.
%DV: Thilina, since we now have space and outline is important with journal papers, I think we should have the outline. But please feel free to remove if more space is needed.

\section{Related work}
\label{sec:related_work}

\dv{
A long line of research has been conducted in privacy preserving record linkage
(PPRL) over the past three decades~\cite{Chr2020,Vat13,Vat17}. 
As surveyed in~\cite{Gko21}, the existing PPRL techniques 
can be categorised into four generations. 

The first generation of PPRL techniques focused on linking records that have the
exact same quasi-identifiers (QIDs) across data sets, supporting only 
exact matching. The second generation considered fuzzy matching of QIDs to 
account for errors and variations in the QIDs across different data sets. 
The third generation of techniques considered scalability of the PPRL process
to linking large databases as an important aspect in addition to the support 
for fuzzy matching~\cite{Kar14,Kar17,Vat17}.  

The fourth generation of techniques focused on data driven technologies by developing
more advanced methods tailored for the different Big Data challenges; velocity,
variety, veracity, and volume. A key challenge of Big Data is to improve
the linkage performance in the presence of data errors by using machine 
learning techniques~\cite{Gko21,dey10}. 

Over the last decade several deep learning based techniques
have been proposed to link entities in databases~\cite{Di19,Koo18,Ebr18,Kas19}. 
Kooli et al.~\cite{Koo18} studied the use of deep Neural Networks (DNN) for linking
records where record pairs are classified into matches or non-matches based on 
word embeddings using a DNN model. 
%\cite{Di19} proposes a methodology to produce 
%explainable interpretations of the output of deep 
%learning models for the entity resolution task.
%
DeepER is another deep learning-based linkage approach, that achieves high 
linkage quality by using recurrent neural networks (RNNs) with long short term 
memory (LSTM) hidden units~\cite{Ebr18}. The aim is to convert each record 
to a distributed representation (i.e., a feature vector), which can effectively 
capture similarities between records. \cite{Kas19} uses deep learning for active 
and transfer learning to reduce the cost of manual labelling required for improving 
the accuracy of linking records. This approach allows to learn a 
transferable model from a high-resource setting to a low-resource one, and to 
further adapt to the target data set, active learning is incorporated that 
carefully selects a few informative examples to fine-tune the transferred model.  

However, these existing deep learning-based linkage techniques are not applicable
to PPRL. To the best of our knowledge, no work has so far considered addressing
the privacy constraints in deep learning-based linkage. Supervised learning
classifiers, such as deep learning techniques, that are trained on sensitive 
data sets (as is the case with most linkage applications), can be vulnerable to
inference attacks, especially membership or attribute inference attacks when 
utilised in PPRL~\cite{Meh20}. 

On other other hand, several works have developed provable privacy-preserving 
data encoding algorithms~\cite{alaggan2012,Sch16,xue2020}. 
%These can be categorised into centralised and federated learning models, where 
%differential private noise is added to the (encoded) input data or to the weights 
%of the models that are trained on the local data, respectively.
%
%With regard to the centralised learning models, 
%Several algorithms have been developed to provide 
For example, $\epsilon$-differentially private Bloom filter encoding algorithms 
have been studied~\cite{alaggan2012,Sch16,xue2020}. %\cite{Sch16,xue2020} studied
%variants of differentially private BF encoding algorithm proposed
%in~\cite{alaggan2012}. Experimental results of these methods show similar 
%privacy-utility trade-off. 
Training models on such Bloom filter encoded data with differential privacy guarantees makes
the models resilient against inference attacks that aim to learn about the 
individuals used in the training data set~\cite{xue2020}.

%Privacy-preserving federated learning, which aims to apply learning tasks 
%by sharing the parameters and weights of locally trained models across 
%multiple parties instead of sharing (encoded) data, has attracted a 
%significant number of research studies in the recent times. The trained deep
%learning models are vulnerable to inference attacks and therefore could 
%still leak information about the training data sets. Differential privacy
%has been used in privacy-preserving federated learning by adding noise 
%to local model weights before sharing with other parties, such that 
%individual parties' data cannot be inferred from the model weights.

Few studies have also investigated the federated learning model
by training supervised learning models with privacy guarantees through adding 
differential privacy noise to the gradients~\cite{Aba16,Nas20}. A successfully 
used method for differential private federated learning is applying
differential privacy in the form of a differentially private stochastic gradient
descent (DP-SGD) optimiser~\cite{Aba16}. DP-SGD can be prohibitively slow to 
train, and hence~\cite{Dup21} proposes an efficient DP-SGD for training using 
a GPU infrastructure.

In our work, we use differentially private Bloom filter encoding with private 
federated learning for applying deep learning-based classification for PPRL. 
To the best of our knowledge, this is the first work to address the challenges 
of applying deep learning for PPRL with improved linkage quality and provable 
privacy guarantees.
}

\section{Background}
\label{sec:background}
We now explain how QID values in a record are encoded  
into a Bloom filter and the application of differential privacy in Bloom filters. 

\begin{figure}[!t]
  \centering
  \includegraphics[width=0.5\textwidth]{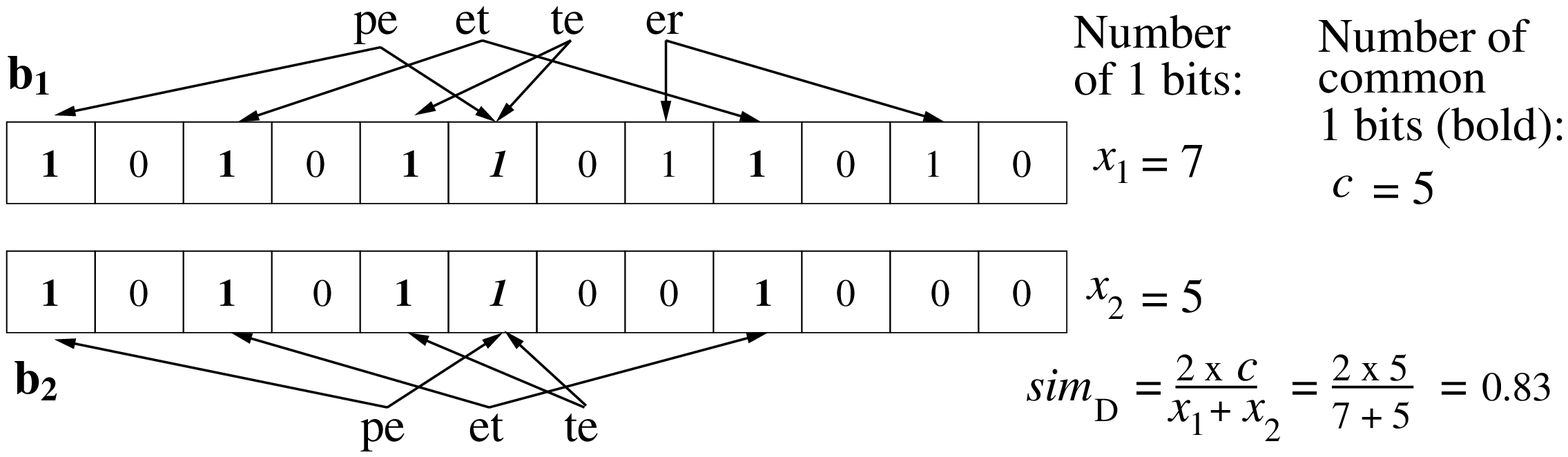}
  \caption{The Dice similarity calculation between the
    names `peter' and `pete', converted into bigrams ($q=2$) and then
    encoded into two Bloom filters $\mathbf{b}_1$ and $\mathbf{b}_2$
    of length $l=12$ bits using $k=2$ hash functions. The bit "$1$" shown
    in italics at position $6$ in $\mathbf{b}_1$ and $\mathbf{b}_2$ is
    a hash collision, because both `pe' and `te' are hashed to this
    position.}
    \label{fig:bloomfilter}
\end{figure}

\subsection{Bloom filters}
\label{sec:bf}

Bloom filter (BF) encoding was proposed by Schnell et al.~\cite{Sch11} for PPRL 
because BFs can be used to efficiently calculate approximate similarities 
between records. 
A BF~\cite{Blo70}  $b$ is a bit vector of length $l= |b|$ where
initially all bits are set to $0$. 
Each data information element in a set $s\in S$ is transformed into
$l$ bits using $k> 1$ hash functions, 
where each hash function outputs an index value between $0$ and $l- 1$. These
index values are then used to toggle the corresponding bits in vector $b$ to $1$. 
In PPRL, 
the set $s$ is generally generated as q-grams, 
i.e., substrings of consecutive characters with a length $q$,  
from one or more QID values from each record in a database, 
as shown in Figure~\ref{fig:bloomfilter}, where various methods
have been proposed to encode strings~\cite{Boy15,Dur14,Pit18,Sch09,vai22,Vat13} 
as well as numerical and sequence data values~\cite{Vat14c,xue2020}. It has however been shown that
BF encoding can be vulnerable to privacy attacks~\cite{Chr17,chr18a, Chr18}. 
As we show in Section~\ref{sec:experiments}, sensitive values that occur 
frequently in an encoded database can lead to frequent bit patterns in BFs 
that can be identified~\cite{Chr2020}, and even individual frequent q-grams 
can be found using pattern mining techniques~\cite{Chr18}.

%\md{[Ming]: Can we provide 1~2 sentences on the privacy vulnerability of the BF encoding?}
% TR: added a sentence to address the above comment. 

\iffalse
\subsection{Deep learning}
% details on deep learning 
% TR: not sure if we need details of deep learning here. I left it 
% just for the moment. 
In the past few years, machine learning-based techniques have become popular in
record linkage applications. Especially, deep learning-based approaches, such as
DeepMatcher, DeepER, and DITTO, use embeddings to represent the entities in a 
database as vector representations which are then used calculated similarities on
the vector representations as features to train a deep learning classifier which
decides if a pair of records is a match or a non-match.

These proposed deep learning-based approaches can capture the features more 
accurately on the entity vector representations which enable them to classify the
record pairs more accurately than traditional Fellengi-Sunter and similarity 
threshold-based record linkage applications. However, these supervised approaches
require larger ground truth data set to train their learning models to achieve an
acceptable level of accuracy. Nevertheless, the training of such models requires
longer runtimes due to iterative training processes consists of these approaches.
\fi

\begin{figure*}[t!]
  \centering
  \includegraphics[width=0.8\textwidth]
  {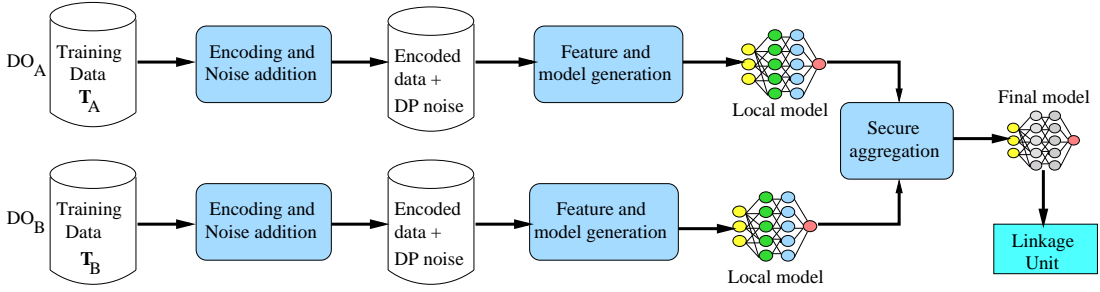}
  \caption{Overview of the training phase of our approach for two database owners 
  (DOs). As we explain in 
  Section~\ref{sec:training}, to ensure privacy each DO adds differentially 
  private noise into their training data in the encoding step. We then utilise
  a secure aggregator in the training step to combine the individual models 
  generated by DOs into a global classification model which will be used in 
  the classification phase as we describe in Section~\ref{sec:classify}.}
  \label{fig:train}
\end{figure*}

\subsection{Differential Privacy}

Differential privacy~\cite{dwork2014} is a privacy definition 
that guarantees the outcome of a calculation to be insensitive to any
particular record in the data set. Differential privacy
requires the output of a data analysis mechanism to be approximately
the same if any single record is replaced with a new one. 
In order to obtain this privacy guarantee, the algorithm employed to 
compute the result of the analysis must contain some form of randomness
such that the probability of obtaining a particular outcome $o\in O$ from 
database $D$ is associated to any pair database-outcome $(D, o)$. Formally:
\begin{definition}[Neighbouring databases]
Databases $\mathbf{D}\in\mathcal{D}$ and $\mathbf{D'}\in\mathcal{D}$ over 
a domain $\mathcal{D}$ are called neighbouring databases if they differ in 
exactly one record.
\end{definition}

\begin{definition}[Differential Privacy~\cite{Dwork2006}]
\label{def:dp}
A randomised algorithm $\mathcal{A}$ is $\epsilon$-differentially 
private if for all neighbouring databases 
$\mathbf{D}$ and $\mathbf{D'}$, and for 
all sets $\mathcal{O}$ outputs, we have
\begin{equation*}
    Pr[\mathcal{A}(\mathbf{D})\in\mathcal{O}]\le
    exp(\epsilon)\cdot Pr[\mathcal{A}(D')\in\mathcal{O}],
\end{equation*}
where $Pr[\cdot]$ denotes the probability of an event.
\end{definition}

In the context of BF, random noise can be added into BFs to guarantee
differential privacy. \dv{The most commonly used method to add random 
noise to BFs for differential privacy guarantees is using the 
randomised response method that  flips certain bit positions 
(from 0 to 1 or from 1 to 0) in each BF with a certain probability.} %at random.
Different mechanisms have been introduced for adding random noise to
BFs~\cite{Sch16,xue2020}. 

Bloom and fliP (known as BLIP) is a method that flips bit values at certain
positions in a BF with a bit flip probability $p$. 
%$p = \frac{1}{1+e^{\epsilon/n}}$ to achieve $\epsilon$-differential 
%privacy~\cite{alaggan2012}, where $n$ is the maximum number of q-grams 
%from a record that are hash-mapped into the BF.  
Formally, 
for a given bit flipping 
probability $p$, a bit $b[i]$ in a BF $b$ at position $i$ is flipped according to:
\begin{equation}
 b[i] = \left\{\begin{array}{lr}
    1~~ ~~~~ \text{if}~ b[i] = 0~
        \text{with probability}~ p, \\
    0~~ ~~~~ \text{if}~ b[i] = 1~
        \text{with probability}~ p, \\
    b[i]~~\, \text{with probability}~ 1 - p.
  \end{array} \right.
  \label{eqn:blip_alaggan}
\end{equation}

Schnell and Borgs~\cite{Sch16} applied the RAPPOR bit flipping method proposed by
Erlingsson et al.~\cite{erlingsson2014}. Assuming again a
flip probability $p$, the bit value $b[i]$ at position
$i$ is flipped according to:
\begin{equation}
  b[i] = \left\{ \begin{array}{lr}
    1~~~~~ ~ \text{with probability}~ \frac{1}{2}p, \\
    0~~~~~ ~ \text{with probability}~ \frac{1}{2}p, \\
    \mathbf{\it{b}}[i]~~ \text{with probability}~ 1 - p. \\
  \end{array} \right.
  \label{eqn:blip_rappor}
\end{equation}

\begin{table}[!t]
    \addtolength{\tabcolsep}{-3.0pt}

  \centering
  \caption{Estimated number of 1-bits in a Bloom filter (BF) of 
    length $l=1000$ bits after applying the two bit flipping approaches
    ~\cite{alaggan2012, erlingsson2014}
    for different fill percentages, FP, (number of 1-bits) in the 
    BF and different flip probabilities, $p$. 
    \label{tab:harden-blip}}
  % \resizebox{0.8\columnwidth}{!}{%
  \begin{small}%\begin{footnotesize} % \begin{scriptsize}
  \begin{tabular}{lccccccccc}  \hline\noalign{\smallskip}
    \textbf{FP (\%)} &
      \multicolumn{3}{c}{~ \textbf{25 (250 1-bits)}} &
      \multicolumn{3}{c}{~ \textbf{50 (500 1-bits)}} &
      \multicolumn{3}{c}{~ \textbf{75 (750 1-bits)}} \\
          	\noalign{\smallskip}\hline\noalign{\smallskip}
   $p$ & 0.01 & 0.05 & 0.1 & 0.01 & 0.05 &
    0.1 & 0.01 & 0.05 & 0.1 \\
	\noalign{\smallskip}\hline\noalign{\smallskip}
  Eq.~\ref{eqn:blip_alaggan}~\cite{alaggan2012} & ~262 &
    ~275 & 300 & ~500 & ~500 & 500 & ~712 & ~725 & 700 \\
  Eq.~\ref{eqn:blip_rappor}~\cite{erlingsson2014}~ &
    ~256 & ~262 & 275 & ~500 & ~500 & 500 & ~725 & ~737 & 725 \\
    \noalign{\smallskip} \hline
  \end{tabular}
 \end{small}%\end{footnotesize} % \end{scriptsize}
\end{table}

For example, if the flip probability is set to $p=0.1$ for a
BF of length $l=$\,1,000 bits, then around 100 bits will
be randomly selected and flipped using Eq.~\ref{eqn:blip_alaggan},
while the rest are unchanged. On the other hand, according to 
Eq.~\ref{eqn:blip_rappor} in~\cite{Sch16} bits are not flipped 
based on their original state in the BF $b$, 
rather around 100 randomly selected bits are either set to $0$ or $1$ 
with equal probability $p/2$. Table~\ref{tab:harden-blip} shows the
estimated number of bits flipped in a BF with different number of 1-bits
for different flip probabilities.

%\md{[Ming]: I think the numbers of bits flip should be stochastic values.
%However, we are showing deterministic values in Table 1.}
% TR: change the sentence to estimated to address the above comment. 

% TR: do we need details about these techniques?
% might be worth to have some details over them, I guess

\section{Deep Learning based PPRL}
\label{sec:method}

%Records in databases are assumed to represent observations of
%entities taken from a particular population.
%Record linkage is defined as the problem of determining which
%records in different databases refer to the same
%entity~\cite{Dong2015morgan,Fel69}. 
Without loss of generality, we assume that two deduplicated 
databases~\cite{Chr12}, \textbf{D}$_A$ and \textbf{D}$_B$, belong
to two database owners (DOs) Alice and Bob, such 
that each entity is represented by only one
record in a database. 
%We assume the same set of quasi-identifies, $A$, is available in 
%both databases, but each database contain different numbers of 
%missing values and errors in each quasi-identifying attribute 
%$a\in A$.
%
%\begin{definition}[Record Linkage]
%
We define two disjoint sets, $\mathbf{M}$ and $\mathbf{U}$, from
the cross-product % of \textbf{D}$_A$ with \textbf{D}$_B$,
\textbf{D}$_A$×\textbf{D}$_B$. 
A record pair $(r_i,r_j)$, with $r_i \in \mathbf{D}_A$ and $r_j \in
\mathbf{D}_B$, is a member $\mathbf{M}$ (true matches) if
$(r_i,r_j)$ represents the same entity $e_i$; otherwise $(r_i,r_j)$ 
is a member of $\mathbf{U}$ (true non-matches) and $r_i$ and $r_j$
represent two different entities $e_i$ and $e_j$, respectively. 
With the true class ($\mathbf{M}$ or $\mathbf{U}$) unknown, the 
privacy-preserving record linkage (PPRL) process attempts to accurately
classify each record pair $(r_i,r_j)$ as belonging to either $\mathbf{M}$ or 
$\mathbf{U}$~\cite{Chr12,Fel69} while preserving privacy of each 
entity $e$ in \textbf{D}$_A$ and \textbf{D}$_B$.

%\md{I think a key question could be on "the motivation of distributed 
%learning in the context of record linkage". If the data record 
%distribution is IID and the data size is sufficient, then I suppose 
%$DO_A$ can train a very good model alone. Otherwise, distributed 
%learning will be beneficial. Is their a way to demonstrate the advantage
%of distributed learning in our experiments?}
%\tr{I think we add such motivation in Introduction. We need to specify
%that one party having large training data is unlikely in real world, but 
%it is likely individual data owners to hold small amount of training data. 
%Thus, we need distributed learning.}

The proposed approach consists of two main phases, known as \emph{training}
and \emph{classification}. Similar to a federated learning setting,
we allow different DOs to train a classification model collaboratively
in the training phase. 
%
%\md{Do we need to specify "different parties" as "the involved two parties A and B"?}
%\tr{I changed the wording to database owners}
%
This classification model is then sent to a %semi-honest
linkage unit (LU) to classify the unlabelled record pairs in the differentially
private BF databases sent to 
it by the DOs into matches and non-matches. We next describe each phase in more details.

%This protocol aims to add differentially
%private noise into encoded training data and train the 
%classification model. The proposed training phase is shown below. 

\subsection{Training phase}
\label{sec:training}

In our approach we assume each DO has a small training data set 
that has been manually labelled before. We assume the record pairs
available in this training data set contain the same quasi-identifiers
as the unlabelled data set allowing us to calculate same set of features 
over the records. The main steps of the training phase is illustrated 
in Figure~\ref{fig:train}.

\begin{figure}[t!]
\begin{center}
  \label{algo:gen-model}
%  \begin{scriptsize}
  \begin{footnotesize}
  \begin{tabular}{ll} \hline

  \noalign{\smallskip}
  \multicolumn{2}{l}{\textbf{Algorithm~1: \emph{Local Model Generation by a DO}}} \\ 
  \noalign{\smallskip} \hline \noalign{\smallskip}
  \multicolumn{2}{l}{Input:} \\
  \multicolumn{2}{l}{- $\mathbf{T}$: \hspace{1mm}\,Training data set
  \hspace{13.5mm}- $q$: \hspace{1.8mm}Q-gram length} \\
  \multicolumn{2}{l}{- $A$: \hspace{1mm} List of QID attributes
  \hspace{8mm}- $l$: \hspace{2mm}Bloom filter length} \\
  \multicolumn{2}{l}{- $H$: \hspace{1.3mm}List of $k$ hash functions 
  \hspace{5mm} - $p$: \hspace{1mm} Bit flip probability} \\
  \multicolumn{2}{l}{- $F$: \hspace{1.8mm}List of feature functions} \\
  \noalign{\smallskip}
  \multicolumn{2}{l}{Output:} \\
  \multicolumn{2}{l}{- $\mathbf{C}$: Local classification Model} \\ \noalign{\smallskip}
  1:  & $\mathbf{C} = [\ ]$, $\mathbf{V} = [\ ]$
        \hspace*{\fill} //  Initialise variables \\
  2:  & \textbf{foreach} $ ((r_i,\ r_j),\ t) \in \mathbf{T}$ \textbf{do:} 
        \hspace*{\fill} //  Loop over training data\\
  3:  & \hspace{2mm} $b_i = \mathit{genBloomFilter}(r_i, A, q, l, H)$ 
  \hspace*{\fill} // Generate BF \\  
  4:  & \hspace{2mm} $b_j = \mathit{genBloomFilter}(r_j, A, q, l, H)$  
  \hspace*{\fill} // Generate BF\\
  5:  & \hspace{2mm} $b'_i = \mathit{addDPNoise}(b_i, p)$  
  \hspace*{\fill} //  Add DP noise to BF\\  
  6:  & \hspace{2mm} $b'_j = \mathit{addDPNoise}(b_j, p)$  
  \hspace*{\fill} //  Add DP noise to BF\\
  7:  & \hspace{2mm} $\mathcal{F} = [\ ]$ \hspace*{\fill} // Initialise feature vector \\  
  8:  & \hspace{2mm} \textbf{foreach} $f \in F$ \textbf{do:} 
        \hspace*{\fill} //  Loop over each function $f$\\
  9:  & \hspace{5mm} $s = f(b'_i, b'_j)$ 
		\hspace*{\fill} //  Compute the feature value\\
  10:  & \hspace{5mm} $\mathcal{F}.add(s)$ 
		\hspace*{\fill} // Add the value to the feature vector \\
  11:  & \hspace{2mm} $\mathbf{V}.add((\mathcal{F},\ t))$
        \hspace*{\fill} //  Add feature vector with label \\
  12: & $\mathbf{C}= \mathit{trainModel(\mathbf{V})}$
        \hspace*{\fill} //  Train the model\\        
  13: & return $\mathbf{C}$ \\
      \hline
%    \vspace{-7mm}
  \end{tabular}
%  \end{scriptsize}
  \end{footnotesize}
\end{center}
\end{figure}

As outlined in Algorithm~1, each DO first encodes the record pairs in its 
training data set into Bloom filters (BFs) (line 2). We assume the training data
consist of record pairs ($r_i$, $r_j$) and their classified labels, $t$,
as match or non-match. As we explained in Section~\ref{sec:bf}, in lines 3 and 4,
each record pair ($r_i$, $r_j$) is encoded into BFs, $b_i$ and $b_j$, respectively,
using the function $\mathit{genBloomFilter()}$.

\begin{figure}[t!]
  \centering
  \includegraphics[width=0.45\textwidth]
  {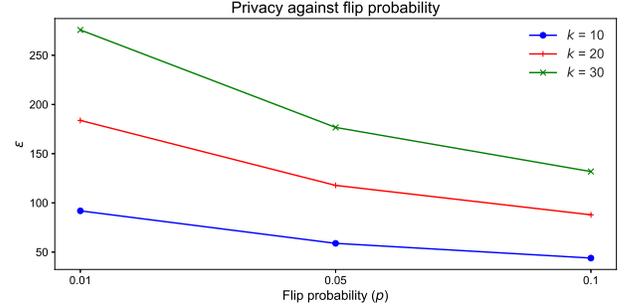}
  \caption{Privacy budget ($\epsilon$) dependent on the bit flipping probability 
           $p$ with $n=10$ for different number of hash functions $k$. 
           Lower $\epsilon$ values denote a higher privacy level.}
  \label{fig:eps-prob-privacy}
\end{figure}

As we explained in Section~\ref{sec:background} and experimentally validate in 
Section~\ref{sec:experiments}, in lines 5 and 6 the function $\mathit{addDPNoise()}$ 
adds differentially private noise into the generated BFs $b_i$ and $b_j$ and 
creates two new BFs $b'_i$ and $b'_j$, respectively. 
Following Definition~\ref{def:dp}, we calculate the bit flip probability, $p$, as,
\begin{equation}
    p= \frac{1}{1+e^{\epsilon /2nk}}
    \label{eq:blip_prob}
\end{equation}
to achieve $\epsilon$-differential 
privacy for a given privacy budget $\epsilon$, 
where $n$ is the maximum number of q-grams from a record that are 
hash-mapped into the BF and $k$ is the number of hash functions. 
%
%\md{If room, I suppose it would be better to present the equation of probability 
%as a formal one with an equation label.}
%
We follow Eq.~\ref{eq:blip_prob} for 
flipping the bits in a BF. 
In Section~\ref{sec:analysis} we theoretically prove that this provides 
$\epsilon$-differential privacy guarantees.
%In Section~\ref{sec:analysis} we analyse the 
%trade-off between privacy and linkage quality in our approach for different 
%$p$ values and 
Fig.~\ref{fig:eps-prob-privacy} shows the corresponding $\epsilon$ 
values for different flip probabilities $p$. It is important to note that, due to
the randomisation in the $addDPNoise$() function, 
the bit values in different positions are flipped in each BF in the pair 
$(b_i, b_j)$ in the bit flipping process.

%\md{Please confirm that we will add different DP noise for different pairs of
%$(b_i, b_j)$. Please note that it means the perturbed $b_i'$ would be 
%different when paired with different $b_j$.}
%\tr{In here, DP noise is added only once to BFs. }

In line 8 we iterate over each feature function $f$ in the list $F$
to generate a feature vector $\mathcal{F}$ for the noise added 
BF pair ($b'i$, $b'_j$). We use different similarity functions,
such as Hamming distance, Dice similarity, Jaccard similarity, etc,
as feature functions where each similarity/distance value computed 
between ($b'i$, $b'_j$) is added as a feature into $\mathcal{F}$. 
Once the feature vector $\mathcal{F}$ is generated, 
we append the corresponding ground-truth label $t$ to $\mathcal{F}$ 
and create the train data list $\mathbf{V}$ (line 11). 

In line 12, each DO will use the list $\mathbf{V}$ to train a deep learning 
classifier $\mathbf{C}$ using the $trainModel$() function. 
Different neural network architectures have 
been proposed to consider different types of embedding structures in 
entity resolution literature,
the most popular of which uses a recurrent structure~\cite{Ebr18}. 
We use unidirectional recurrent neural networks (RNN) with long 
short term memory (LSTM) hidden units~\cite{hochreiter1997}, 
thus called as LSTM-RNNs. In our approach, all DOs agree on the 
parameter settings to be used on the local models. We provide the details 
of the hyper-parameter selection of individual local models in 
Section~\ref{sec:experiments}.
Finally, once the local models $\mathbf{C}$ are generated by all the DOs
the local models are aggregated into a final global model $\mathbf{C_G}$ 
as we illustrate in Fig.~\ref{fig:train}.
%
%\md{Figure~\ref{fig:train} might be a bit misleading since it only 
%illustrates two parties. We might need to clarify that the number of 
%DOs could be larger than 2 in practical applications.}
%\tr{In the beginning of Section 4, we mentioned we only assume
%two parties,without loss of generality. But I guess we can change
%the figure if needed.}

Following a centralised federated learning 
setting~\cite{kairouz2021}, we utilise a secure aggregator (SA) in this 
phase to combine the local models sent by the DOs. The use of SA does not
allow the LU to learn information about the local models as we explain in
Section~\ref{sec:analysis}. Following~\cite{mcmahan2017}, SA uses a 
federated averaging approach to take the average of all weights sent by
DOs. The average of the weights is regarded as the new set of weights 
for the global model. The final global model $\mathbf{C_G}$ is then 
sent to the LU to be used in the classification phase as we describe next. 

%\md{We might need to explain the purpose of SA a bit more. SA is not for
%privacy, am I right? I think privacy has been taken care of by the DP 
%noise. Is SA almost the same as a dedicated server?}
%\tr{I think SA does provide privacy from the LU from learning anything
%about the local models. SA is similar to a server which is commonly used
%in the FL context.}
% add few sentences about the computational complexity of the 
% algorithm

\begin{figure*}[!t]
  \centering
  \includegraphics[width=0.8\textwidth]
  {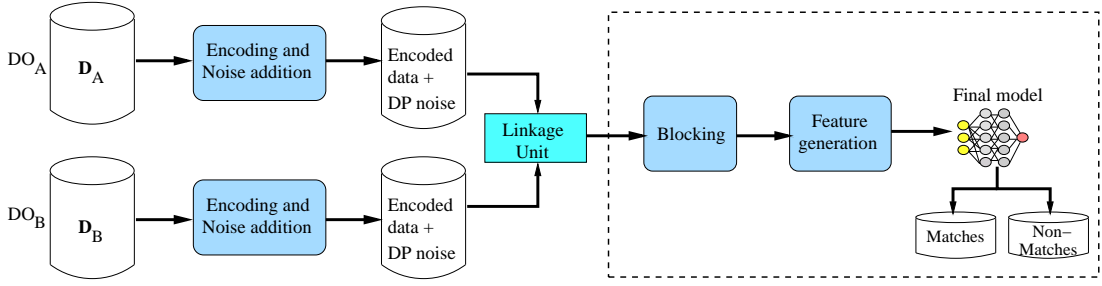}
  \caption{Overview of the classification phase with two DOs.
           The steps within the dashed box are performed by the 
           LU. }
  \label{fig:classify}
\end{figure*}

\subsection{Classification phase}
\label{sec:classify}

%This main steps of the training phase is illustrated in Fig~\ref{fig:test}. 
As illustrated in Fig.~\ref{fig:classify}, 
the DOs first encode their databases following the same parameter setting used 
in the training phase. 
%
%\md{I think Fig.~\ref{fig:classify} is misleading in the sense that the 
%"encoded" database is "encoded" and noise-perturbed in the first step. 
%Should be remove the superscript "e" in $D_A$ and $D_B$?}
%\tr{Good point, I will update the figure.}
%
Each DO then adds differentially private noise to their encoded 
database $\mathbf{D^e}$. As in Algorithm~1, the DOs use the functions 
$\mathit{genBloomFilter()}$ and $\mathit{addDPNoise()}$ to encode 
each attribute value in the list of attributes $A$ of each record 
in their database into BFs and add differentially private noise to
these generated BFs, respectively. The noise-perturbed encoded databases 
are then sent to the LU.

\begin{figure}[t!]
\begin{center}
  \label{algo:classify}
%  \begin{scriptsize}
  \begin{footnotesize}
  \begin{tabular}{ll} \hline \noalign{\smallskip}
  \multicolumn{2}{l}{\textbf{Algorithm~2: \emph{Classification by the LU}}} \\ 
  \noalign{\smallskip} \hline \noalign{\smallskip}
  \multicolumn{2}{l}{Input:} \\
  \multicolumn{2}{l}{- $\mathbf{D_A^e}$: \hspace{1mm}\,Alice's encoded database
  \hspace{11mm}- $\mathbf{C_G}$: \hspace{1mm} The global model}\\
  \multicolumn{2}{l}{- $\mathbf{D_B^e}$: \hspace{1mm} Bob's encoded database
  \hspace{12.5mm}- $F$: \hspace{4.2mm}%List of 
  Feature functions} \\
  %\multicolumn{2}{l}{- $\mathbf{C_G}$: \hspace{1mm} The global model} \\
  %\multicolumn{2}{l}{- $F$: \hspace{4.2mm}List of feature functions} \\
  \noalign{\smallskip}
  \multicolumn{2}{l}{Output:} \\
  \multicolumn{2}{l}{- $\mathbf{M}$: Classified match record pais} \\ \noalign{\smallskip}
  1:  & $\mathbf{M} = [\ ]$, $\mathbf{B} = \{\ \}$
        \hspace*{\fill} //  Initialise variables \\
  2:  & $\mathbf{B}$ = $genBlocks(\mathbf{D_A^e}, \mathbf{D_B^e})$ 
        \hspace*{\fill} //  Generate blocks \\        
  3:  & \textbf{foreach} $ (r_i.id,\ r_j.id,\ b'_i,\ b'_j) \in \mathbf{B}$ \textbf{do:} 
        \hspace*{\fill} // Loop over blocks \\
  4:  & \hspace{2mm} $\mathcal{F} = [\ ]$ \hspace*{\fill} // Initialise feature vector \\  
  5:  & \hspace{2mm} \textbf{foreach} $f \in F$ \textbf{do:} 
        \hspace*{\fill} //  Loop over each function $f$\\
  6:  & \hspace{5mm} $s = f(b'_i, b'_j)$ 
		\hspace*{\fill} //  Compute the feature value\\
  7:  & \hspace{5mm} $\mathcal{F}.add(s)$ 
        \hspace*{\fill} // Add the value to the feature vector \\
  8:  & \hspace{2mm} $\mathbf{m}= \mathit{classify(\mathbf{C_G}, \mathcal{F})}$
        \hspace*{\fill} //  Classify the record pair\\
  9:  & \hspace{2mm} \textbf{if} $m == match$ \textbf{do:} \\        
  10:  & \hspace{5mm} $\mathbf{M}.add((r_i.id, r_j.id))$ 
  \hspace*{\fill} // Add the record pair to matches\\
  11: & return $\mathbf{M}$ \\
      \hline
%    \vspace{-7mm}
  \end{tabular}
%  \end{scriptsize}
  \end{footnotesize}
\end{center}
\end{figure}

We outline the classification process for two encoded databases
in Algorithm~2. In line 2 the LU first applies a blocking 
technique~\cite{kar15} to the encoded databases to generate
candidate record pairs. We assume this blocking technique to be a 
\emph{black box} as any appropriate private blocking technique can 
be used~\cite{Chr2020,Dur12,kar15}. The LU then iterates over each candidate
record pair and applies the feature generation step by using the same list of
feature functions $F$ used in the training phase (lines 5 to 7). 

Next, in line 8, each feature vector $\mathcal{F}$ is then classified 
using the global model $C_G$ it received from the secure aggregator to 
classify each candidate record pair as a match or non-match. If the 
candidate record pair is classified as a match then the corresponding 
record identifier pair $(r_i.id, r_j.id)$ is added to the set of matches 
$\mathbf{M}$ (line 10). Once all the candidate record pairs are classified, 
the LU sends $\mathbf{M}$ to the DOs (line 11).

%the candidate record pairs where each candidate pair results in a 
%feature vector. These feature vectors are then passed into the final 
%classification model it received from a secure aggregator to classify 
%each candidate record pair as a match or non-match.  
%Once the classification model is generated it is sent to the linkage
%unit to be used to classify unlabelled data. 

% add few sentences about the computational complexity of the 
% algorithm

\section{Privacy Analysis}
\label{sec:analysis}

%We now analyse our approach with regard to privacy, linkage quality, and scalability.

%\subsection{Privacy analysis}

\dv{
Two possible methods can be used for preserving privacy in the 
training phase: 1) adding differential privacy noise to the encoded
Bloom filters (BFs) and training the local models on the noise 
added BFs to make the trained local models robust against 
inference attacks when shared, or 2) adding differential privacy noise 
to the trained model weights and exchanging only the noise added model
weights. 

In the context of PPRL, 
the linkage of unlabelled records requires exchanging or sharing the (encoded) 
records between the DOs or with a third party (linkage unit), 
and thus the (encoded) records need to be privatised in any case.
Hence, in our approach the method (1) is more appropriate, 
because for linking or classifying record pairs using the trained 
model, the similarities between records from different DOs need to 
be computed and used by the LU. 

Moreover, training the model using training data and classifying unlabelled 
record pairs need to be conducted on the same feature space, i.e. 
training the model on non-perturbed BFs and classifying the 
perturbed BFs of unlabelled record pairs, would not learn correctly 
and perform accurate linkage.
Thus, we apply differential privacy
noise to the BFs and train the deep learning model, 
which is then used by the LU to classify the (differentially private)
BFs of the unlabelled record pairs. 

Specifically, the BFs are perturbed by using the randomised response
method that flips the bits with a probability of 
$\frac{1}{1+e^{\epsilon/2nk}}$ to achieve 
$\epsilon$-differential privacy, where $k$ is the number of hash functions 
used to hash-map tokens (e.g. $q$-grams for string attributes or neighbouring
values for numerical attributes) into the BFs, and $n$ is the 
maximum possible number of tokens in any record in a data set (assuming 
$n \ll l$, where $l$ is the length of the BFs).

\begin{theorem}
Flipping the bits in BFs with 
$\frac{1}{1+e^{\epsilon/2nk}}$ probability, where $k$
is the number of hash functions and $n$ is the maximum possible number 
of tokens in any record in a data set, makes the BFs 
$\epsilon$-differentially private.
\end{theorem}
}
%The detailed proof is presented in Appendix A.

% need to get the statistics out of these data sets
\begin{table*}[t!]
\centering
  \caption{Overview of the data sets used in the experimental evaluation. 
  \label{tab:datasets}}
  \begin{small}
\begin{tabular}{lcccccc}
  \noalign{\smallskip}\hline\noalign{\smallskip}
  Data set & Domain & Num Records & Ground Truth & Training data & Missing values (\%) & Provenance \\
  \noalign{\smallskip}\hline\noalign{\smallskip}
  DBLP-ACM (Clean)      & Publication & 2,616 - 2,294 & 2,224 & 7,417 & 0 - 0 & Real\\
  DBLP-ACM (Dirty)      & Publication & 2,616 - 2,294 & 2,224 & 7,417 & 49 - 47 & Real\\  
  DBLP-Scholar (Clean)  & Publication & 2,616 - 64,263 & 2,324 & 17,223 & 7 - 22 & Real\\
  DBLP-Scholar (Dirty)  & Publication & 2,616 - 64,263 & 2,324 & 17,223 & 48 - 61 & Real\\  
 \noalign{\smallskip}\hline\noalign{\smallskip}
  iTune-Amazon  & Music  & 6,907 - 55,923 & 132 & 321 & 0 - 0 & Real\\
  Music-Brainz  & Music  & 3,827 - 3,861 & 1,219 & 3,565 & 24 - 23 & Real\\
  \noalign{\smallskip}\hline\noalign{\smallskip}
   Amazon-Google & E-commerce & 1,363 - 3,226 & 1,300 & 3,571 & 8 - 10 & Real\\
  \noalign{\smallskip}\hline\noalign{\smallskip}
  NCVR                  & Demographic & 222,251 - 224,061 & 148,036 & 425,731 & 3 - 4  & Real\\
  European Census       & Demographic &  25,343 - 24,613 & 24,043 & 71,887 & 3 - 3 & Synthetic \\
  \noalign{\smallskip}\hline\noalign{\smallskip}
\end{tabular}
\end{small}
\end{table*}

\begin{proof}
Let us assume two Bloom filters (BFs) of any two records $r$ and $r'$
can at maximum differ by $n \times k$ bit positions, when there are no
collision between bits during hash-mapping, where $n$ is the maximum
number of tokens in any record and $k$ is the number of hash functions.

Let $v$ and $v'$ be the BFs produced by records $r$ and $r'$ 
respectively before noise is applied. Let 
$\mathcal{A}: \{0, 1\}^l \rightarrow \{0, 1\}^l$ be a random noise function
such that $\mathcal{A}(i) = i$ with probability 
$\frac{e^{\epsilon/2nk}}{1+e^{\epsilon/2nk}}$, and 
$\mathcal{A}(i) = 1-i$ with probability 
%(i.e. we flip bits with the desired probability) 
$\frac{1}{1+e^{\epsilon/2nk}}$, where $i \in \{0,1\}$.

%This gives us the expression
\begin{equation}
    \frac{Pr[{BF}(r, \epsilon) = \tilde{v}]}{Pr[{BF}(r', \epsilon) = \tilde{v}]} = \prod_{b=1}^l \frac{Pr[\mathcal{A}(v_b) = \tilde{v_b}]}{Pr[\mathcal{A}(v_b') = \tilde{v_b}]}
\label{eq:2}
\end{equation}

Note that the two BFs $v, v' \in \{0,1\}^l$ can only differ in at 
most $2 \times n \times k$ positions, corresponding to $v_{h_i(r_j)} = 1$ 
for all $1 \le i \le k$ and $1 \le j \le n$, and 
$v_t = 0$ for all $t \neq v_{h_i(r_j)}$, while 
$v'_{h_i(r'_j)} = 1$ for all $1 \le i \le k$ and $1 \le j \le n$, and
$v'_t = 0$ for all $t \neq v'_{h_i(r'_j)}$. We can then simplify the
ratio in~(\ref{eq:2}) by considering only the corresponding $n \times k$ 
positions, because all terms $v_t, v'_t = 0$ for 
$t \neq \{v_{h_i(r_j)}, v'_{h_i(r'_j)} \}$.

This ratio is maximised when all $n \times k$ bit positions are different 
in both BFs, and therefore $2nk$ bits in either 
$v'$ or $v$ need to be flipped (maximum ratio).

\begin{equation}
%\begin{multlined}
    e^{-\epsilon} \le 
    \frac{Pr[{BF}(r, \epsilon) = \tilde{v}]}
    {Pr[{BF}(r', \epsilon) = \tilde{v}]} \le \prod_{b=1}^{2nk} \frac{\frac{e^{\epsilon/2nk}}{1+e^{\epsilon/2nk}}}{\frac{1}{1+e^{\epsilon/2nk}}}
    \le e^\epsilon
%\end{multlined}
\label{eq:3}
\end{equation}

Bounding the above ratio, we get
\begin{equation}
    -\epsilon \leq \text{ln} \left(\frac{Pr[{BF}(r, \epsilon) = 
    \tilde{v}]}{Pr[{BF}(r', \epsilon) = \tilde{v}]} \right) \leq \epsilon
\end{equation}
\end{proof}

\section{Experiments}
\label{sec:experiments}
% details of experimental results goes here
We conducted experiments to validate the effectiveness of the proposed approach. 
We first describe the experimental setup, 
including the data sets, baselines and parameter setting, 
and then we discuss in detail the results that we have obtained.
%from these experiments.

\subsection{Experimental setup}

%We first describe the data sets we used in our experiments. We 
%evaluate our system on several data sets in the deep-learning based
%entity resolution (ER) literature. All are popular benchmark data sets
%and have been extensively evaluated by prior ER work using both machine
%learning (ML) and non-ML based approaches. 

\subsubsection{Data sets}

Table~\ref{tab:datasets} shows the details of the data sets used in 
the experiments. 
DBLP-ACM and DBLP-Scholar consist of links between academic 
publications~\cite{Ebr18}. We used the attributes authors names, 
publication name, venue, and year as the linkage attributes. For these two 
data sets, we use two variations Clean and Dirty for experiments, 
where Clean data sets contain records without any data quality issues, 
while Dirty data sets embody various data quality problems, 
such as missing values, misspellings, and variations of values~\cite{Ebr18}.

iTune-Amazon data set contains music data from iTunes and Amazon.
We used the attributes song name, artist name, album name, genre, and 
price. The Music Brainz data set is based on real 
records about songs from the MusicBrainz database~\cite{leipzigdata}. 
We used the attributes song title, artist name, album name, and year in 
the linkage process.
% need to add description of music brainz data set

We used the North Carolina Voter Registration (NCVR) 
database (http://dl.ncsbe.gov/data/) with one snapshot from April 2014 
and a second one from June 2014. We extracted pairs of records that 
correspond to the same voter but had name and/or address changes 
overtime.
%, resulting in 222,251 and 224,061 records, respectively. 
%We used different combinations of the attributes first name, 
%last name, street address, and city in our experiments.
%
We used a synthetic European census database (\url{https://ec.europa.eu/eurostat/cros/content/job-training\_en}) 
generated to represent real observations of the decennial census.
The database contains personal details of fictitious people.
We used the attributes first name, last name, street address, 
and city for both NCVR and European census databases.
%in the linkage process.

Each data set is split into the training, validation, and test sets using
a ratio of 3:1:1. The same split of the data sets is also used in the
evaluation of ER literature~\cite{Ebr18}. For each of the data sets,
we performed K-fold cross validation with $K=5$. 
We report the average results along with the standard deviation. 
%obtained across all the folds.

\begin{table}[t!]
\centering
  \caption{Different feature generation functions. 
  \label{tab:functions}}
\begin{tabular}{l|l}
\hline
\multicolumn{1}{c|}{Similarity functions} & 
\multicolumn{1}{c}{Distance functions}\\ \hline
 Jaccard\cite{Chr2020}, Dice\cite{Chr2020},  &
 Hamming~\cite{Dur12}, Bray-Curtis\cite{choi2010},         \\ 
 Cosine\cite{Vat13}, Russell-Rao\cite{choi2010}, &
 Jensen-Shannon\cite{choi2010},          \\ 
 Yule\cite{choi2010}, Sokal-Sneath\cite{choi2010}, &
 Kulsinski\cite{choi2010}, Minkowski\cite{choi2010},\\
 Sokal-Michener\cite{choi2010}, & Squared Euclidean\cite{choi2010} \\
 Rogers-Tanimoto\cite{choi2010} & Weighted Minkowski\cite{choi2010} \\ \hline
 
\end{tabular}
\end{table}

\subsubsection{Performance evaluation metrics}

We evaluated scalability using runtime, and linkage quality using 
precision, recall, and F-measure~\cite{Chr2020}. We consider 
$TP$, $FP$, and $FN$ as the number of true matches, false matches,
and false non-matches, respectively~\cite{Chr12}. 
Precision (P) can be computed as $\text{P} = TP/(TP+FP)$ which measures the
number of true matched record pairs against the total number of matched 
record pairs generated by a particular approach; and recall (R) can be 
computed as $\text{R} = TP/ (TP + FN)$ which measures the number of true matched
record pairs against the total number of record pairs in the linked ground
truth data~\cite{Han18}. F-measure (F) can be computed as 
$\text{F}=(2*P*R)/ (P + R)$ which provides the harmonic mean between 
precision and recall. 

Further, we also used a novel measure called F*-measure in our evaluation
because recent research has shown that F-measure is not suitable for 
measuring linkage quality in record linkage due to the relative importance 
given to precision and recall, which depends upon the number of predicted
matches~\cite{Han18,Han21}. The $F^*$-measure (F*) is 
calculated as $\text{F}^* = \frac{TP}{(TP + FP + FN)}$, which corresponds
to the number of true matches against the number of matches which are either 
misclassified or are correctly classified.
%The details of the definition of F*-measure is provided in Appendix B.

\begin{table*}[t!]
\centering
  \caption{Average runtime results (in seconds) for linking the
  different data sets.
  \label{tab:runtime}}
  \begin{small}
\begin{tabular}{lccccccccc}
\hline\noalign{\smallskip}
Method   & 
\begin{tabular}[c]{@{}c@{}}DBLP-\\ ACM-C\end{tabular}~ & 
\begin{tabular}[c]{@{}c@{}}DBLP-\\ ACM-D\end{tabular}~ & 
\begin{tabular}[c]{@{}c@{}}DBLP-\\ Scholar-C\end{tabular}~ & \begin{tabular}[c]{@{}c@{}}DBLP-\\ Scholar-D\end{tabular}~ & \begin{tabular}[c]{@{}c@{}}iTune-\\ Amazon\end{tabular}~ & \begin{tabular}[c]{@{}c@{}}Music-\\ Brainz\end{tabular}~ & \begin{tabular}[c]{@{}c@{}}Amazon-\\ Google\end{tabular}~ & 
NCVR~ & 
\begin{tabular}[c]{@{}c@{}}European\\ Census\end{tabular}~ \\
\hline\noalign{\smallskip}
BF~\cite{ant19,Ran14,Vat17}  & 6 & 6 & 70 & 71 & 91 & 16 & 11 & 462 & 58 \\
\hline\noalign{\smallskip}
DP-BF~\cite{alaggan2012,Sch16,xue2020} & 8 & 9 & 75 & 75 & 94 & 18 & 13 & 473 & 73 \\
\hline\noalign{\smallskip}
DL-BF & 8 & 9 & 77 & 78 & 96 & 18 & 14 & 475 & 75 \\
\hline\noalign{\smallskip}
DP-DL-BF & 10 & 12 & 92 & 91 & 103 & 25 & 19 & 491 & 82 \\
\hline\noalign{\smallskip}
\end{tabular}
\end{small}
\end{table*}

\begin{table*}[t!]
\centering
  \caption{Average linkage quality results (P: Precision, R: Recall, F: F-measure,
    and F*: F*-measure) of different methods on different data sets. 
    The best F-measure results are shown in bold. 
  \label{tab:linkage-quality}}
\begin{small}
%\begin{footnotesize}
\begin{tabular}{lcccc}
\hline\noalign{\smallskip}
Data set & 
\multicolumn{1}{c}{\begin{tabular}[c]{@{}c@{}}BF\\ P / R / F / F*\end{tabular}} &
\multicolumn{1}{c}{\begin{tabular}[c]{@{}c@{}}DP-BF\\ P / R / F / F*\end{tabular}} &
\multicolumn{1}{c}{\begin{tabular}[c]{@{}c@{}}DL-BF\\ P / R / F / F*\end{tabular}} &
\multicolumn{1}{c}{\begin{tabular}[c]{@{}c@{}}DP-DL-BF\\ P / R / F / F*\end{tabular}} \\ 
\hline\noalign{\smallskip}
DBLP-ACM-C & 
0.94 / 0.80 / 0.82 / 0.69 & 
0.90 / 0.67 / 0.77 / 0.62 & 
0.96 / 0.82 / \textbf{0.88} / 0.79 & 
0.94 / 0.79 / 0.86 / 0.75 \\ 
\hline\noalign{\smallskip}

DBLP-ACM-D & 
0.85 / 0.56 / 0.68 / 0.51 & 
0.82 / 0.45 / 0.58 / 0.41 & 
0.96 / 0.72 / \textbf{0.82} / 0.70 & 
0.93 / 0.73 / \textbf{0.82} / 0.69 \\ 
\hline\noalign{\smallskip}

DBLP-Scholar-C &
0.88 / 0.47 / 0.61 / 0.44 & 
0.59 / 0.34 / 0.43 / 0.28 & 
0.98 / 0.65 / \textbf{0.78} / 0.64 & 
0.99 / 0.43 / 0.61 / 0.44 \\ 
\hline\noalign{\smallskip}

DBLP-Scholar-D &
0.58 / 0.31 / 0.40 / 0.25 & 
0.57 / 0.25 / 0.34 / 0.21 &
0.98 / 0.58 / \textbf{0.72} / 0.57 & 
0.99 / 0.54 / 0.69 / 0.53 \\ 
\hline\noalign{\smallskip}

iTune-Amazon & 
0.68 / 0.43 / 0.52 / 0.35 & 
0.55 / 0.36 / 0.43 / 0.27 & 
0.82 / 0.59 / \textbf{0.68} / 0.52 & 
0.81 / 0.50 / 0.62 / 0.45 \\ 
\hline\noalign{\smallskip}

Music-Brainz & 
0.90 / 0.89 / 0.89 / 0.80 & 
0.96 / 0.51 / 0.66 / 0.49 & 
0.99 / 0.84 / \textbf{0.90} / 0.83 & 
0.99 / 0.72 / 0.83 / 0.71 \\ 
\hline\noalign{\smallskip}

Amazon-Google & 
0.68 / 0.66 / 0.67 / 0.50 & 
0.55 / 0.46 / 0.50 / 0.33 & 
0.85 / 0.69 / \textbf{0.76} / 0.61 & 
0.85 / 0.63 / 0.72 / 0.57 \\ 
\hline\noalign{\smallskip}

NCVR & 
0.89 / 0.69 / 0.78 / 0.63 & 
0.91 / 0.42 / 0.57 / 0.40 & 
0.86 / 0.84 / \textbf{0.84} / 0.74 & 
0.80 / 0.81 / 0.80 / 0.67 \\ 
\hline\noalign{\smallskip}

European Census & 
0.82 / 0.66 / 0.73 / 0.57 & 
0.80 / 0.49 / 0.60 / 0.43 & 
0.99 / 0.82 / \textbf{0.89} / 0.81 & 
0.96 / 0.82 / \textbf{0.89} / 0.79 \\ 
\hline\noalign{\smallskip}
\end{tabular}
%\end{footnotesize}
\end{small}
\end{table*}

% Comparison methods
As baselines we compared our approach (referred as \textbf{DP-DL-BF}) with 
three PPRL techniques. (1) We used a BF based PPRL technique (named as \textbf{BF}), 
which uses a similarity threshold to classify record pairs encoded into BFs as matches and non-matches. 
%However, BF does not provide provable 
%privacy guarantees. 
(2) For the second baseline approach (named as \textbf{DP-BF}), 
we adapted the first baseline approach by adding differentially private noise
into BFs before the linkage. (3) We applied the same deep learning technique
in the classification step of the first baseline approach \textbf{BF} to make the third
baseline, named as \textbf{DL-BF}. In this approach, we did not include any
differentially private noise into BF encodings.

\subsubsection{Parameter values}

Based on a parameter sensitivity analysis, 
we set the similarity threshold to $0.7$ for BF and DP-BF. 
For DL-BF and DP-DL-BF approaches, 
we used three hidden layers with 21, 42, and 84 neurons, respectively. 
We adopted an Adam optimiser with a learning rate of $0.002$. 
We conducted 50 epochs with batch size of 5 in the training step. 
Table~\ref{tab:functions} shows the similarity and distance functions we used as the list of feature functions, ${F}$, 
in Algorithms~1 and 2, 
where these functions are commonly used in PPRL and pattern matching for binary data~\cite{choi2010,Chr2020,Dur12,Vat13}. 
Following~\cite{erlingsson2014,Sch16}, 
we set bit flip probability $p=[0.01, 0.05, 0.1]$ in Algorithms~1 and 2 to ensure the same amount of DP noise is added in BFs and set the number of hash functions $k=[10, 20, 30]$. 
We used the q-gram length $q=2$, 
the BF length $l=1,000$ bits, 
and double hashing~\cite{Chr2020} to encode q-grams into BFs. 
We adopted a Hamming distance based locality sensitive hashing 
technique~\cite{Dur12,Kar14} for blocking.

\subsubsection{Experiment environments}

We implemented all approaches in Python 3.7. We used TensorFlow to
implement our deep learning model~\cite{abad16}. 
All experiments were performed on a 64-bit Intel Core i9 chip, 
with eight cores running 16 threads at speeds of up to 2.4GHz, 
along with 64 GBytes of memory, 
and running Windows 10. 
To facilitate repeatability, 
the data sets and the programs are available from the authors. 
% we might need to upload then somewhere - public access

\subsection{Results and discussion}

\subsubsection{Comparison of runtime}

Table~\ref{tab:runtime} shows the runtime of all approaches on different data sets. 
As expected, 
our approach consumes more runtime compared to other approaches because of the addition of 
differentially private noise in BFs and the independent training of the local models by each DO. 
DP-BF and DL-BF need a similar runtime due to the independent noise addition and local model training by each DO, respectively, 
as classification of unlabelled record pairs by the LU requires a small amount of additional runtime for 
feature generation for the record pairs generated by the blocking technique. 
Thus, 
the use of an efficient blocking technique can reduce the additional overhead of runtime required by the LU to a minimum.

\begin{figure*}[!t]
  \centering
%  \includegraphics[width=0.33\textwidth]
%  {figures/reident-prob-tableA_clean_2.eps}
%  \includegraphics[width=0.33\textwidth]
%  {figures/reident-prob-tableA_clean_1-2.eps}
%  \includegraphics[width=0.33\textwidth]
%  {figures/reident-prob-tableA_clean_1-2-3-4.eps}
%  \caption{Re-identification results for a frequency based 
%  attack~\cite{Chr17} for different flip probability ($p$) for different
%  attribute combinations with the DBLP-ACM (Clean) data set.}
%  \label{fig:re-ident-prob-acm}
%  \end{figure*}
%  \begin{figure*}[!t]
%  \centering

%  \caption{Re-identification results for a frequency based attack~\cite{Chr17}
% for different flip probability ($p$) for different attribute combinations with
%the NCVR data set.}

%\end{figure*}
%
%\begin{figure*}[!t]
%  \centering
  \includegraphics[width=0.33\textwidth]
  {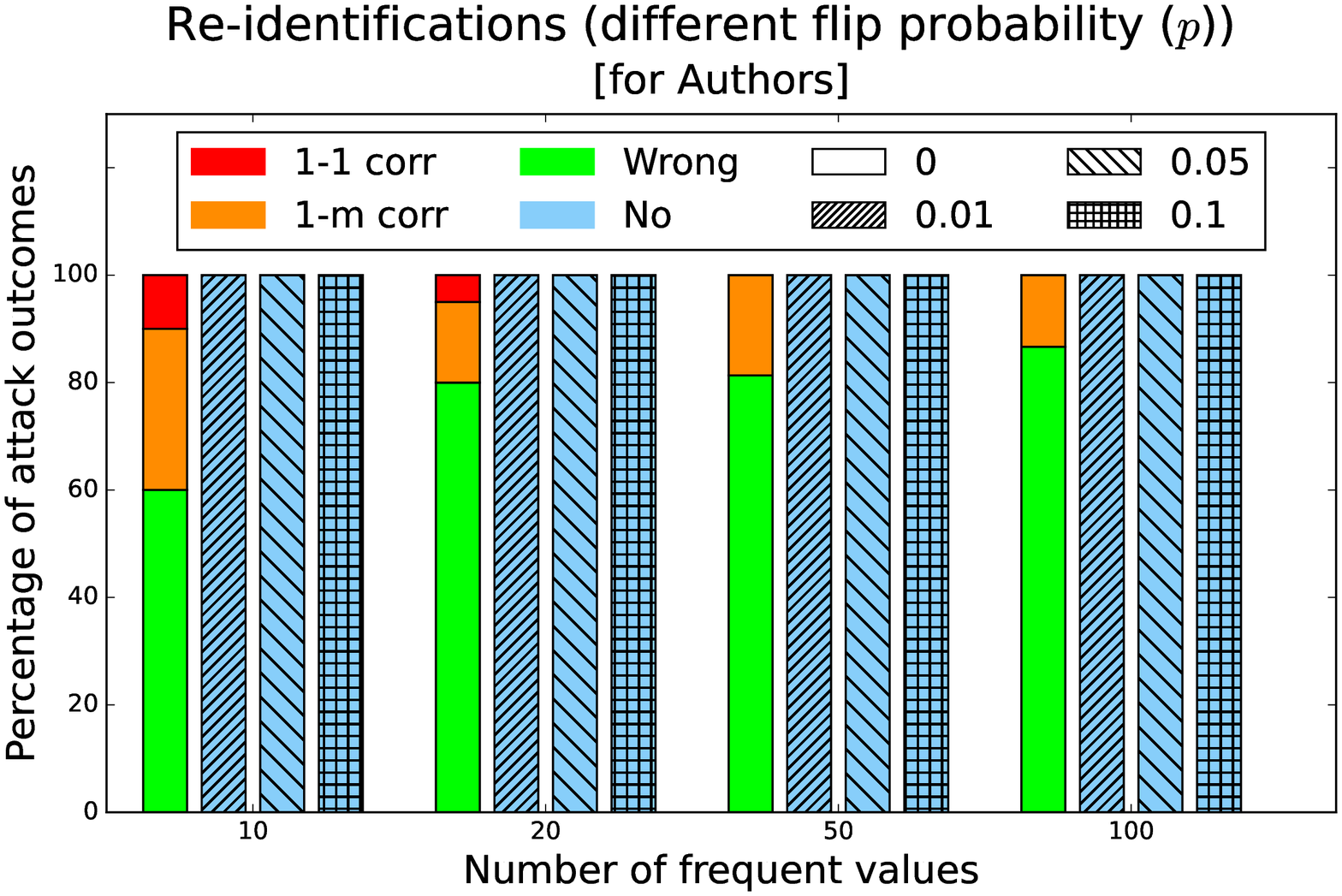}
  \includegraphics[width=0.33\textwidth]
  {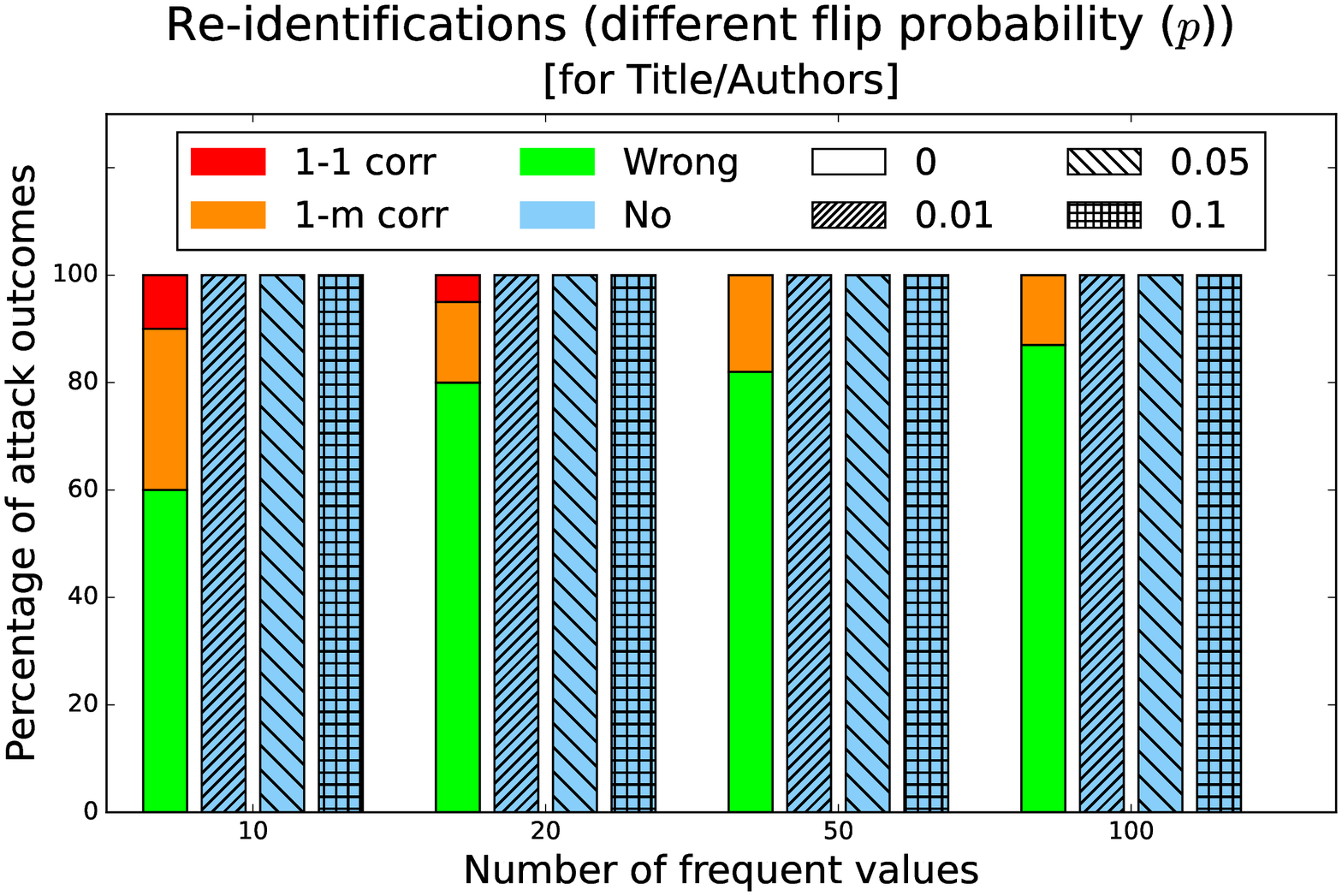}
  \includegraphics[width=0.33\textwidth]
  {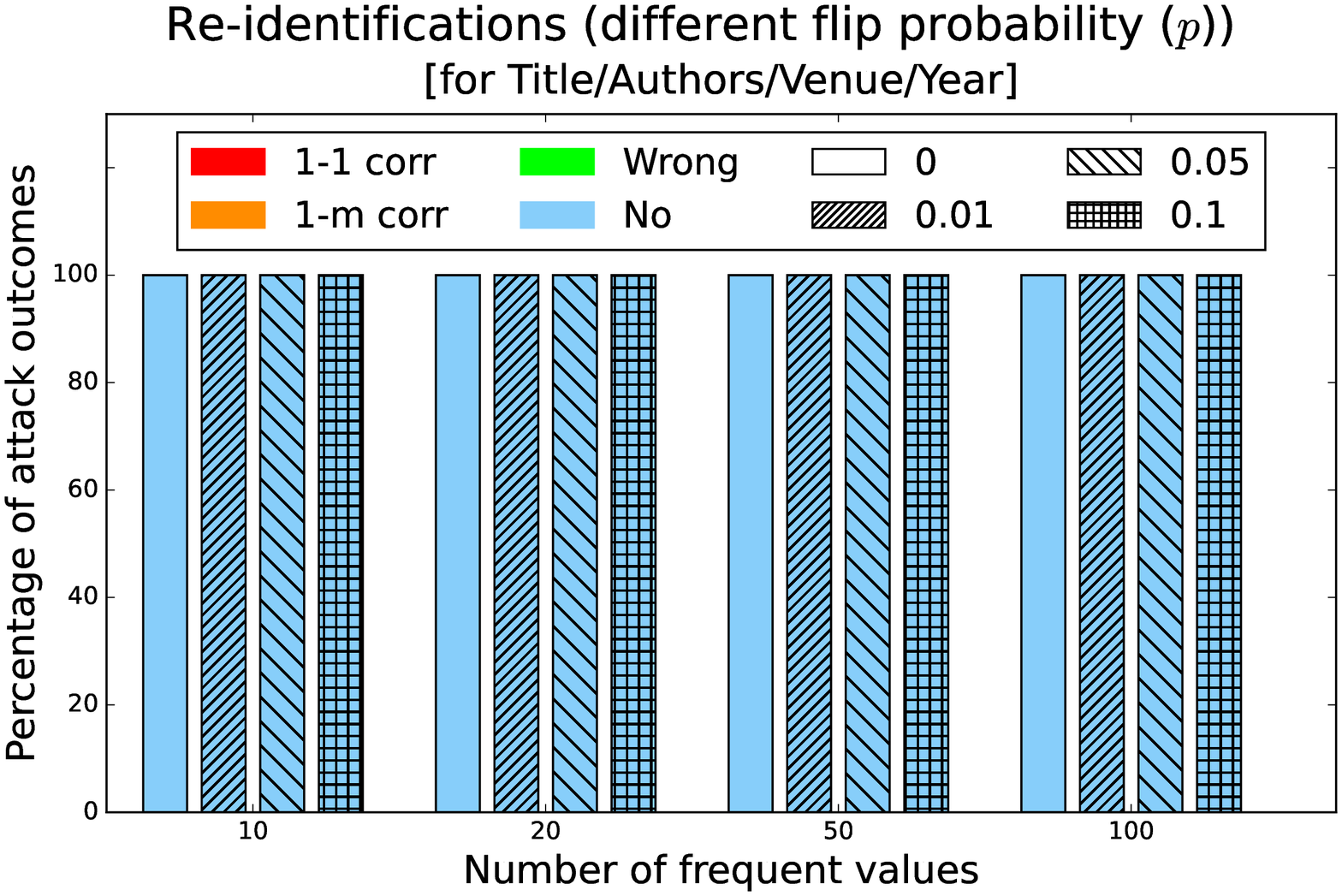}
  \smallskip
    \includegraphics[width=0.33\textwidth]
  {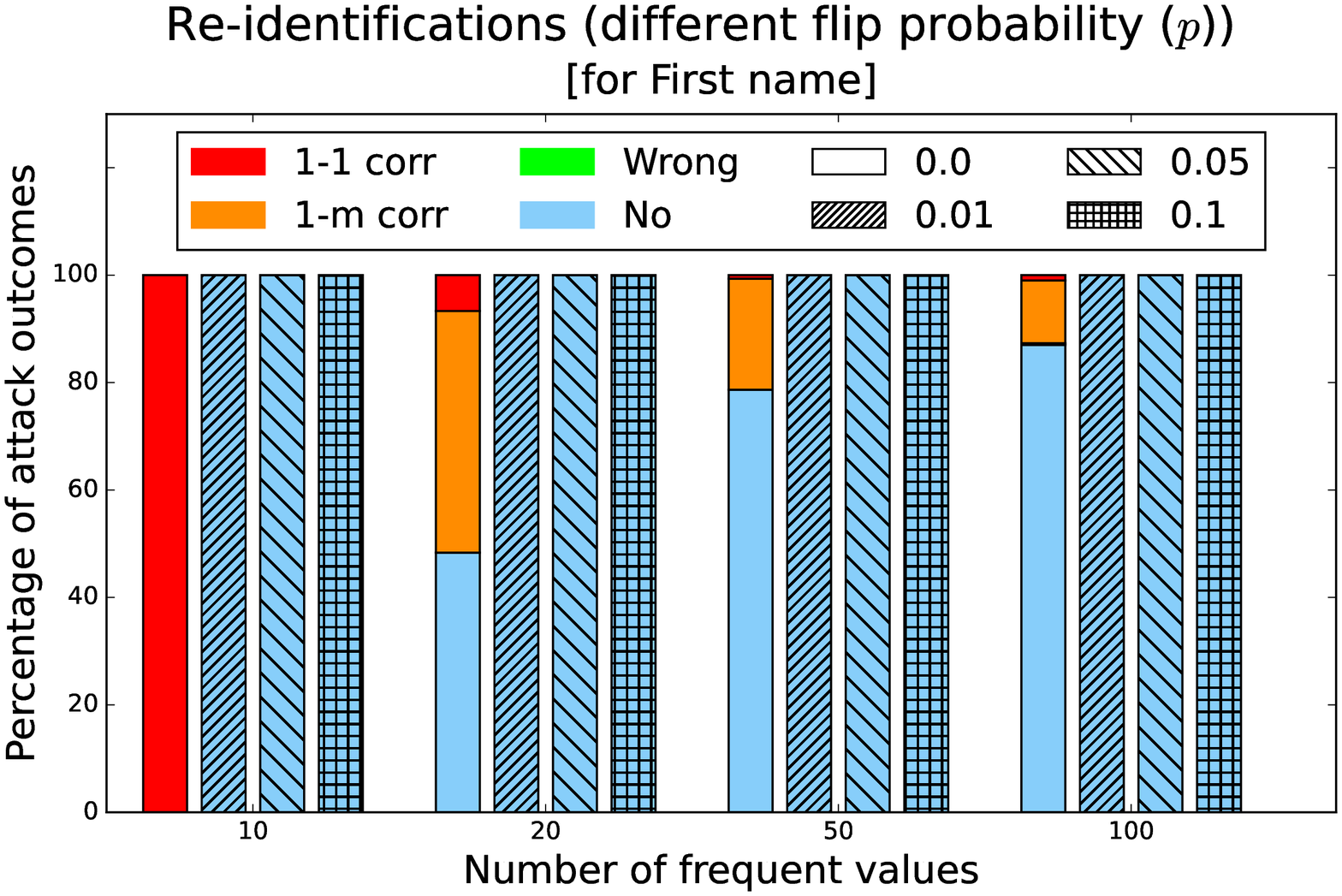}
  \includegraphics[width=0.33\textwidth]
  {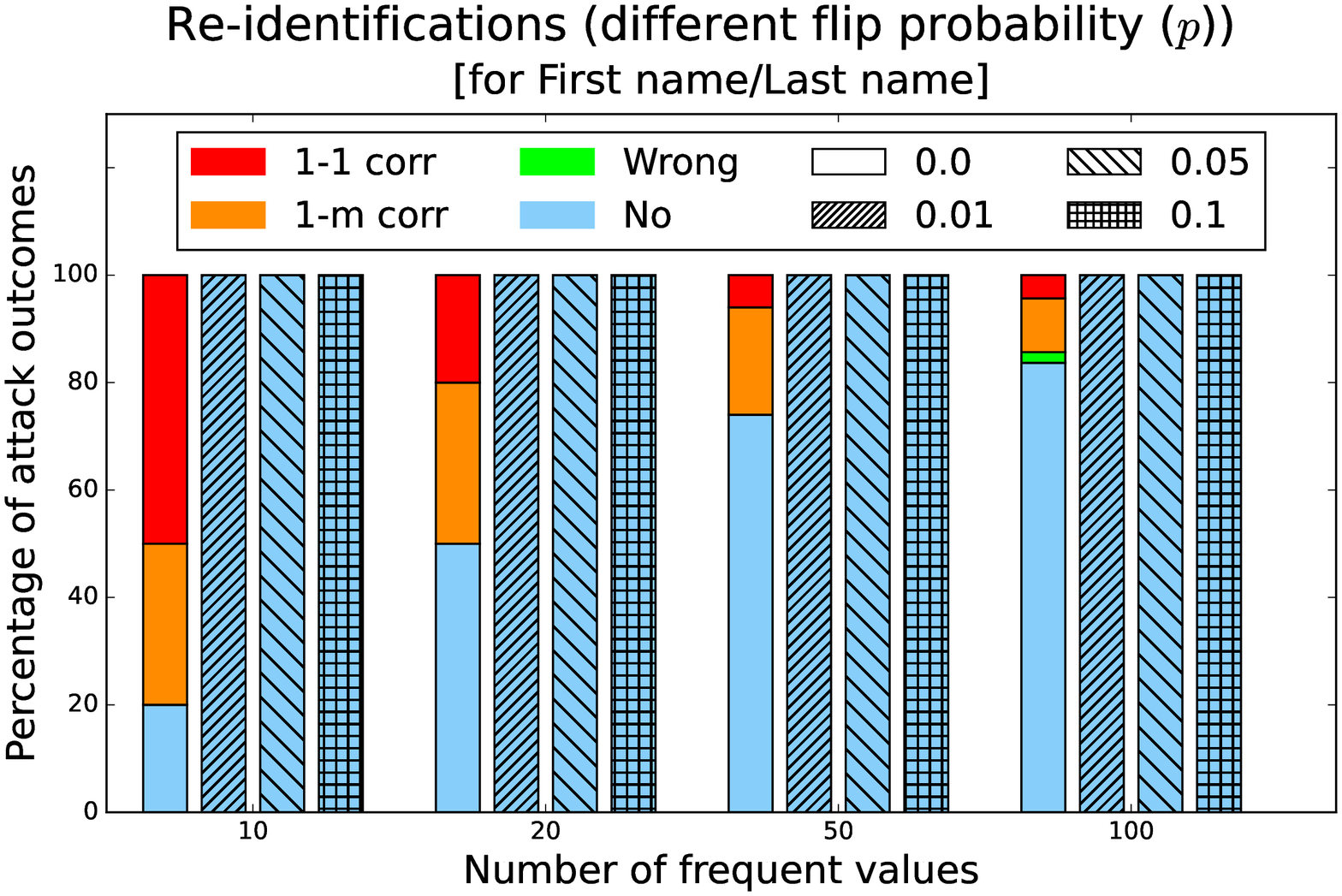}
  \includegraphics[width=0.33\textwidth]
  {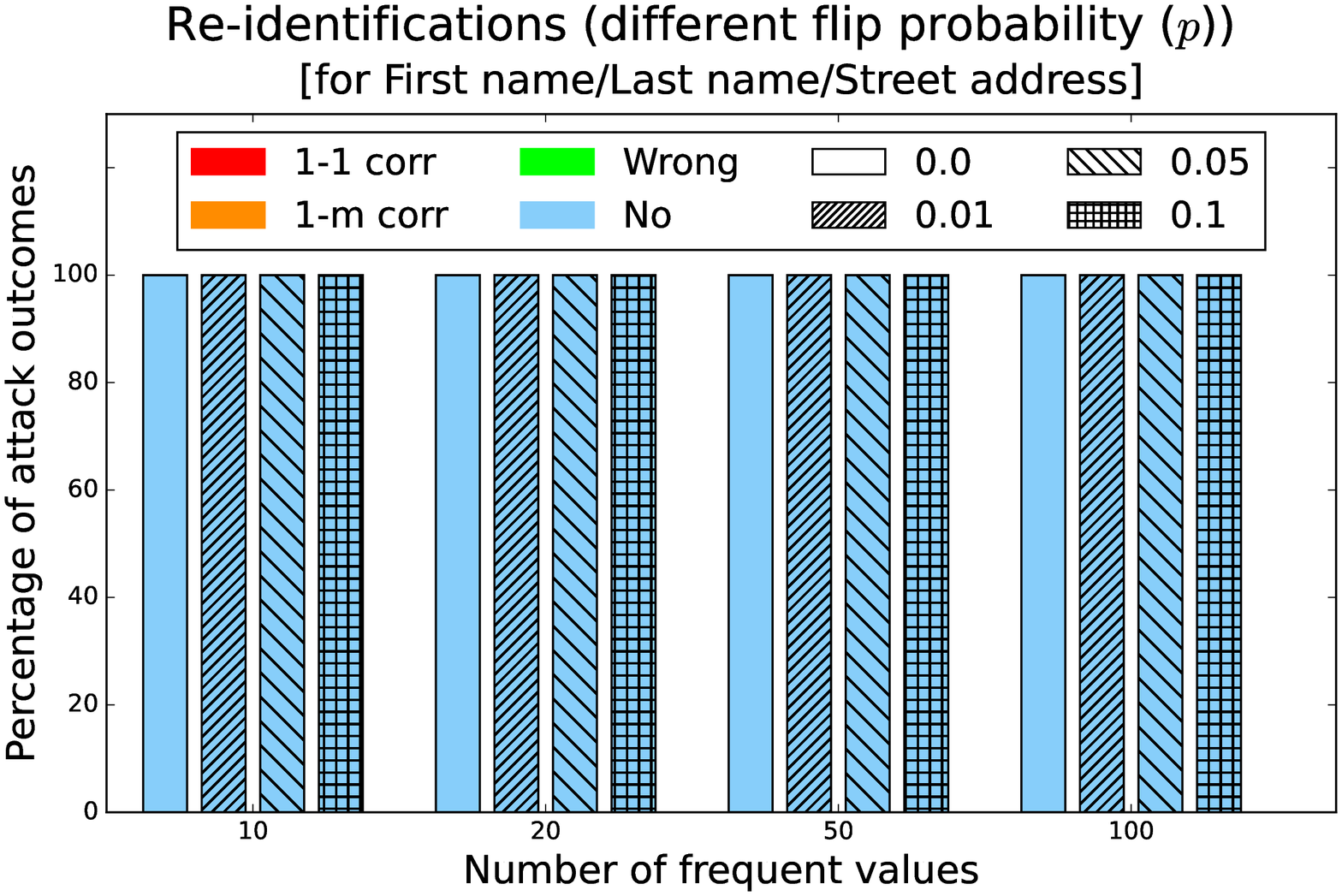}
  \smallskip
  \includegraphics[width=0.33\textwidth]
  {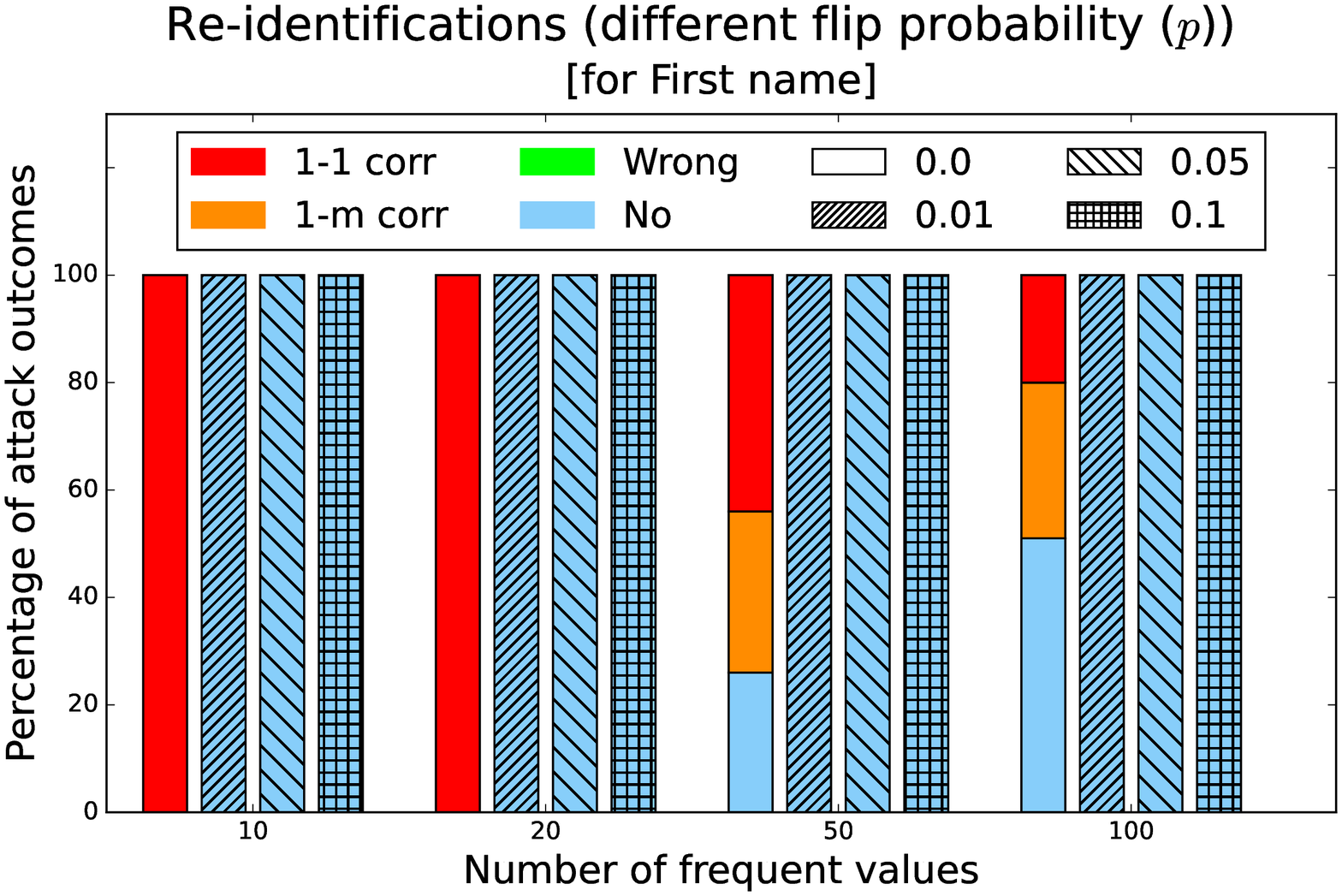}
  \includegraphics[width=0.33\textwidth]
  {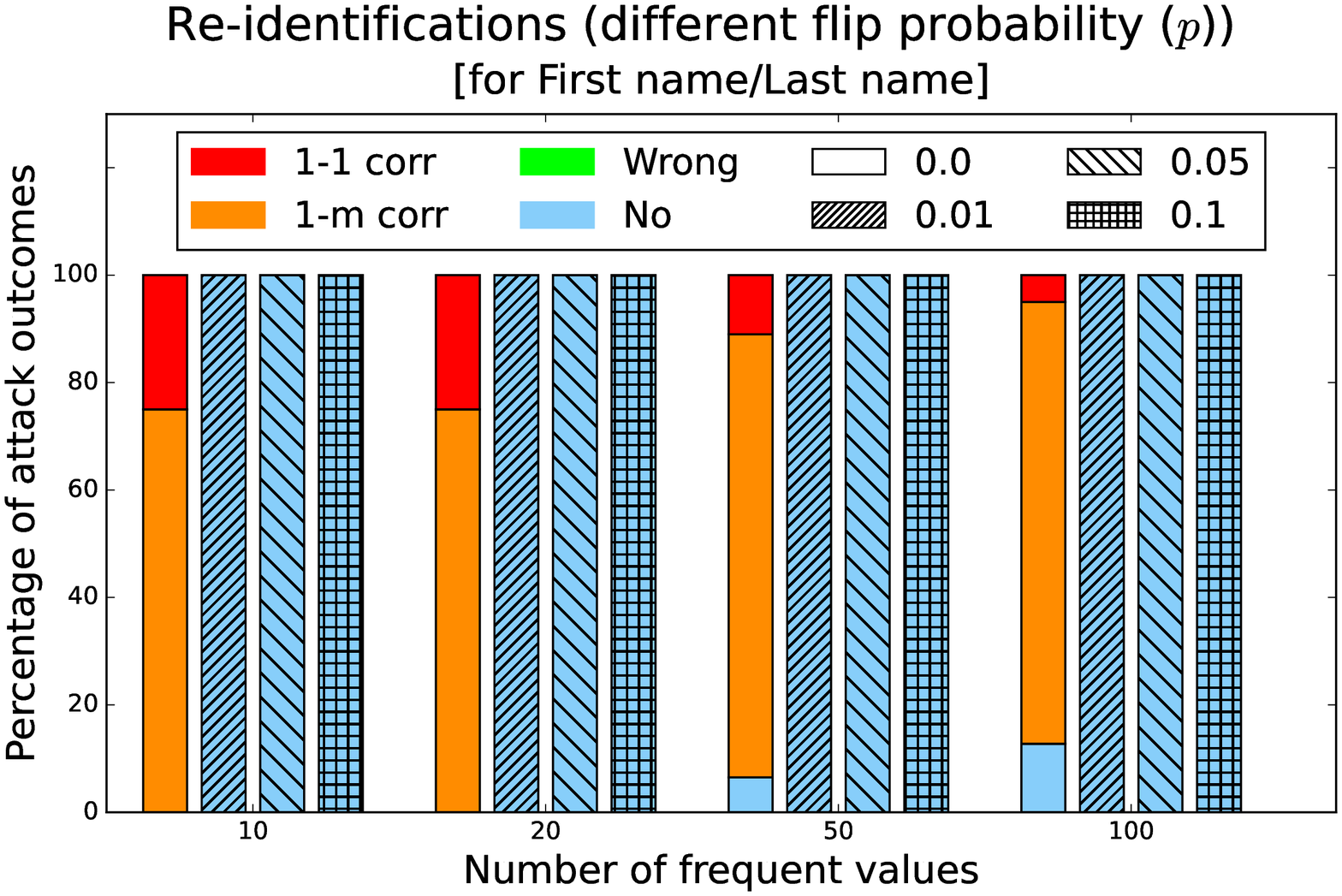}
  \includegraphics[width=0.33\textwidth]
  {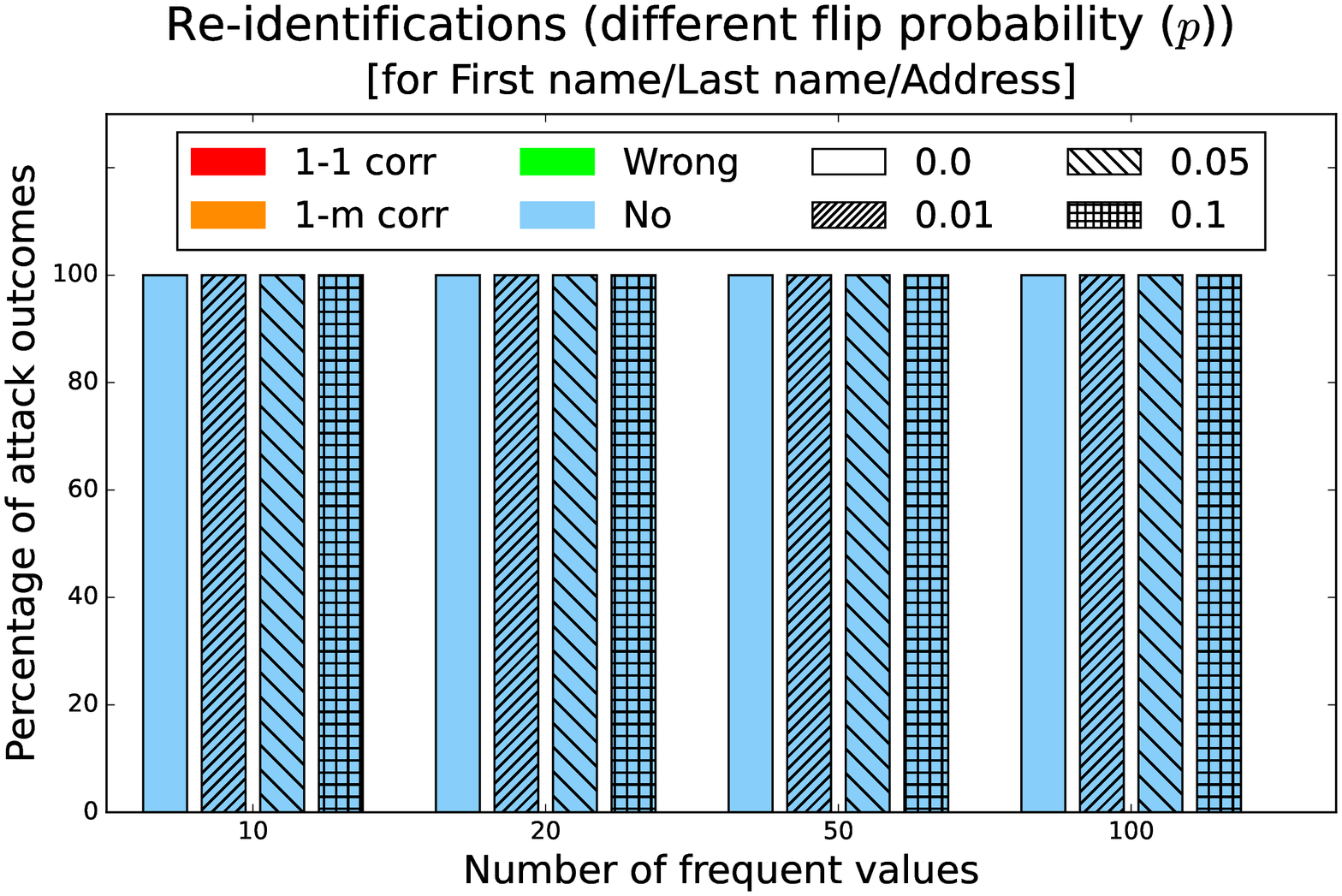}
   \caption{Re-identification results for a frequency based 
  attack~\cite{Chr17,chr18a} for different flip probability ($p$) and for different
  attribute combinations with the DBLP-ACM (Clean) data set (top row),
  the NCVR data set (middle row), and
  European Census data set (bottom row).}
  %\label{fig:re-ident-prob-acm}
  \label{fig:re-ident-prob}
  \end{figure*}

\subsubsection{Linkage quality}

Table~\ref{tab:linkage-quality} shows the linkage quality of all 
approaches on different data sets. As can be seen, DL-BF achieves 
the best precision and recall values among all approaches. 
Our approach, DP-DL-BF achieves a similar linkage quality as DL-BF, 
but, showing a slightly lower recall.
This is because of the noise addition in BFs, 
which leads to true match record pairs being classified as non-matches in the classification phase. However, 
both BF and DP-BF resulted in lower linkage quality (20\% -- 40\% in terms 
of F- and F*-measures) compared to the deep learning based approaches, 
indicating that deep learning models can classify encoded record pairs
with a higher accuracy than the na\"ive threshold based classification technique. 

Further, 
the results in Table~\ref{tab:linkage-quality} show that the addition of differential
privacy noise in BFs can render a threshold based classifier less effective as 
precision and recall drop by at most 33\% and 43\%, respectively. 
This indicates that the use of differential privacy in practical linkage projects 
with threshold based classifiers requires careful fine-tuning of BF encoding and 
threshold parameters to achieve an acceptable level of linkage quality.

\subsubsection{Resistance to privacy attacks}

To evaluate the privacy we used the frequency based cryptanalysis attack by 
Christen et al.~\cite{Chr17,chr18a}. This attack aligns frequent BFs and plain text values 
in a public database to conduct re-identification of the most frequent values encoded
in these BFs. We assume this attack is launched by the LU in the classification phase
on the BFs that the LU receives from all DOs. 

We conducted this attack assuming one file in a data set pair is the encoded BF database
while the other one represents the public database. 
We evaluate the re-identification accuracy in terms of the percentages of 
(1) correct guesses with 1-to-1 matching (1-1 corr), 
(2) correct guesses with 1-to-many (1-m corr) matching, 
(3) wrong guesses (Wrong), 
and (4) no guesses (No), 
where these four percentages sum to $100$. 
We considered the re-identification accuracy of the attack based on identifying 
the 10, 20, 50, and 100 most frequent plain text attribute values from the public
data set~\cite{Chr17,chr18a}.

\begin{figure*}[!t]
  \centering
    \includegraphics[width=0.33\textwidth]
  {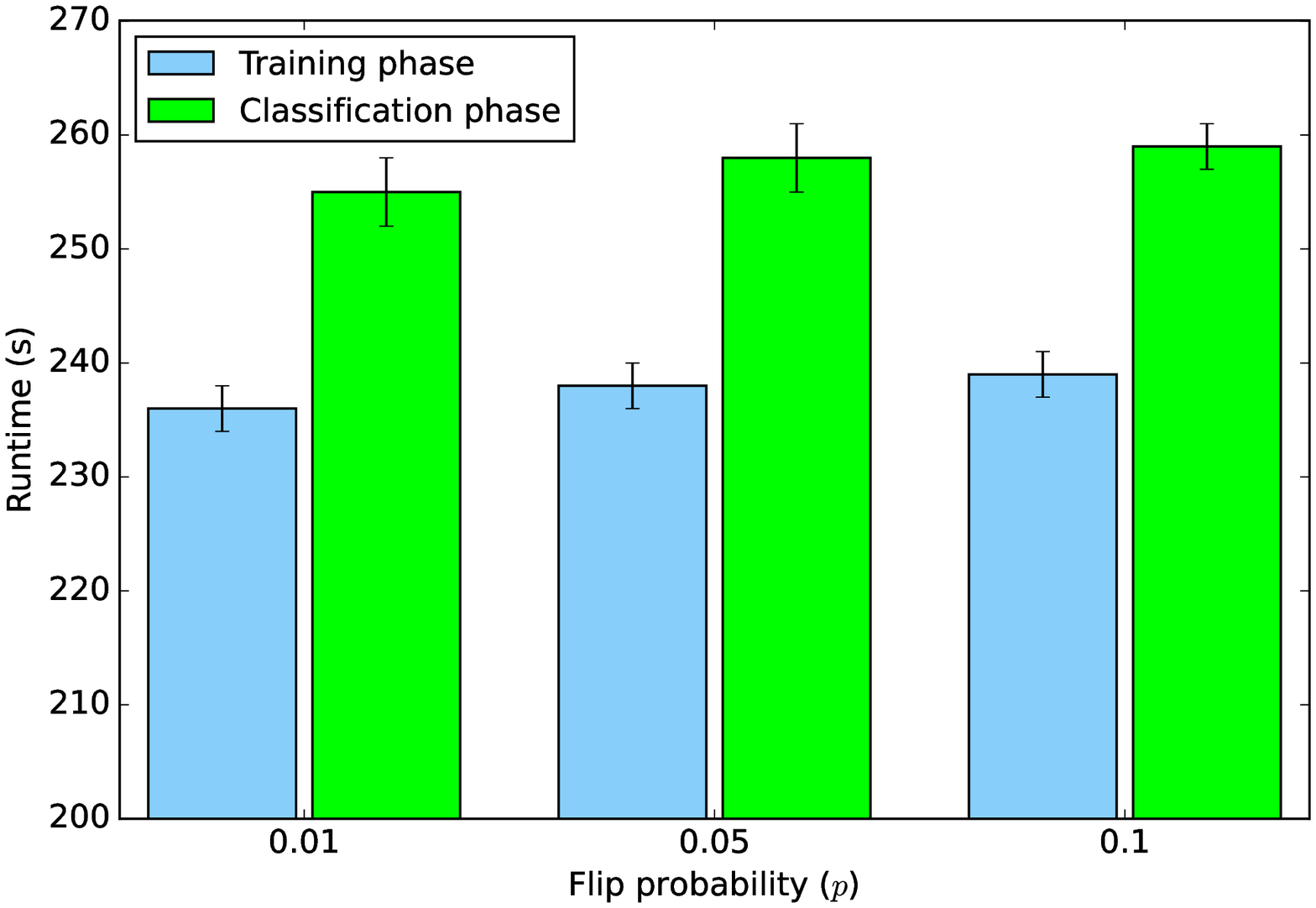}
  \includegraphics[width=0.33\textwidth]
  {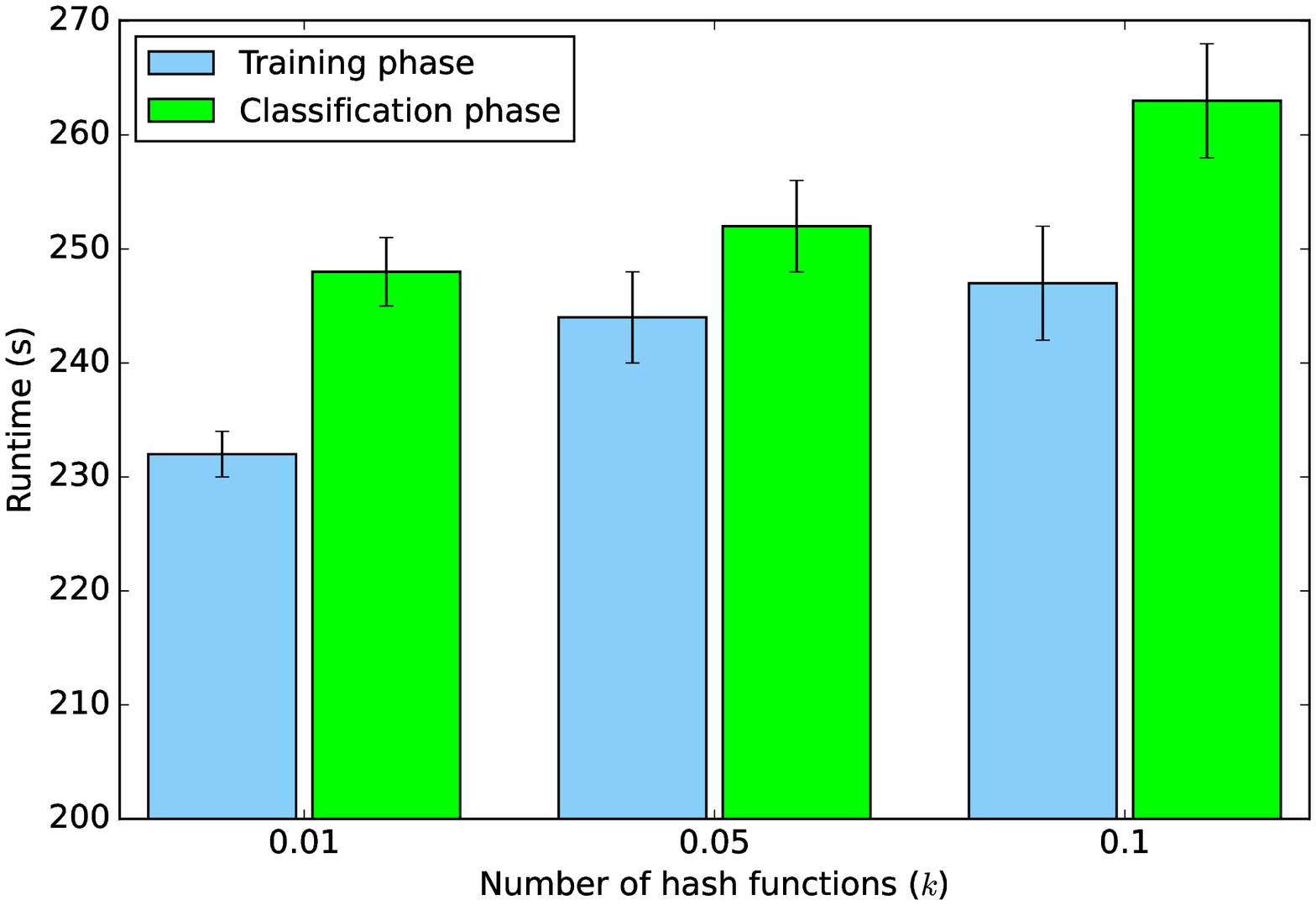}
  \includegraphics[width=0.33\textwidth]
  {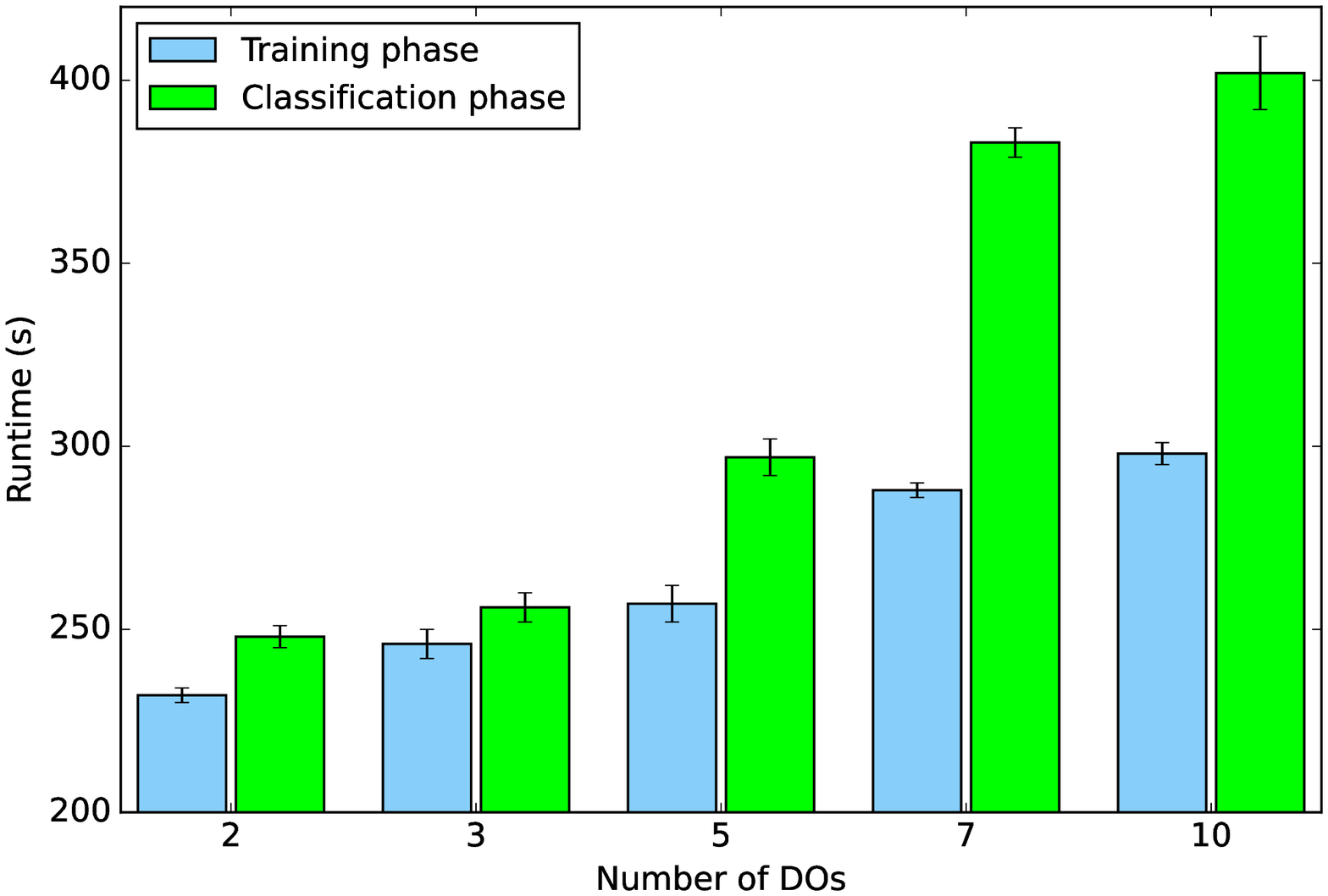}
  \caption{Average runtime results (in seconds) for the NCVR data set.
    Note the different y-axis scales.}
  \label{fig:aba-runtime}
\end{figure*}

Figure~\ref{fig:re-ident-prob} shows the re-identification results for the
DBLP-ACM-C (top row), NCVR (middle row), and European Census (bottom row) data sets. 
As can be seen from the figure, without DP noise ($p$ = 0) in BFs 
(as in BF and DL-BF), the attack identified frequent attribute values correctly
for different attribute combinations.
For the DBLP-ACM-C data set the attack correctly identified the 4 most frequent 
author names correctly. For NCVR, 
the attack correctly identified the 10 most frequent first names 
%while with the DBLP-ACM-C data set the attack identified 3 most frequent
%author names 
correctly while for the European Census data set the attack 
re-identified the top 20 first names correctly.
Further, as our privacy results show, the attack failed to correctly 
identify the attribute values encoded in BFs when bits are flipped 
according to DP noise ($p>0$). This is because these encoded BFs contain 
different bit patterns, leading the cryptanalysis attack unable to 
identify any frequent BFs. 

\begin{table*}[t!]
    \addtolength{\tabcolsep}{-4.0pt}
\centering
  \caption{Average linkage quality results (P: Precision, R: Recall,
  F: F-measure, and F*: F*-measure) with different flip 
  probability ($p$) on different data sets. 
  \label{tab:linkage-quality-p}}
%\begin{footnotesize}
\begin{small}
\begin{tabular}{llll}
\hline
\multirow{2}{*}{Data set} & 
\multicolumn{3}{c}{Flip probability ($p$)}
\\ \cline{2-4}
 & 
 \multicolumn{1}{c}{\begin{tabular}[c]{@{}c@{}}0.01 \\ P / R / F / F*\end{tabular}}&
 \multicolumn{1}{c}{\begin{tabular}[c]{@{}c@{}}0.05 \\ P / R / F / F*\end{tabular}} &
 \multicolumn{1}{c}{\begin{tabular}[c]{@{}c@{}}0.1 \\ P / R / F / F*\end{tabular}} \\ 
 \hline\noalign{\smallskip}
DBLP-ACM-C &
0.94 / 0.79 / 0.86 / 0.75 ~~~~&
0.86 / 0.75 / 0.80 / 0.67 ~~~~&
0.80 / 0.72 / 0.75 / 0.61 \\ 
\hline\noalign{\smallskip}
DBLP-ACM-D & 
0.93 / 0.73 / 0.82 / 0.69 &
0.90 / 0.67 / 0.76 / 0.62 &
0.90 / 0.65 / 0.75 / 0.61 \\
\hline\noalign{\smallskip}
DBLP-Scholar-C &
0.98 / 0.54 / 0.70 / 0.53 &
0.98 / 0.50 / 0.66 / 0.49 &
0.98 / 0.45 / 0.62 / 0.45 \\ 
\hline\noalign{\smallskip}
DBLP-Scholar-D & 
0.98 / 0.43 / 0.59 / 0.43 &
0.97 / 0.40 / 0.57 / 0.40 &
0.96 / 0.38 / 0.54 / 0.37 \\ 
\hline\noalign{\smallskip}
iTune-Amazon & 
0.81 / 0.50 / 0.62 / 0.45 &
0.82 / 0.49 / 0.61 / 0.44 &
0.81 / 0.35 / 0.49 / 0.32 \\
\hline\noalign{\smallskip}
Music-Brainz & 
0.99 / 0.72 / 0.83 / 0.71 &
0.99 / 0.70 / 0.82 / 0.69 &
0.99 / 0.65 / 0.78 / 0.64 \\
\hline\noalign{\smallskip}
Amazon-Google &
0.85 / 0.63 / 0.72 / 0.57 &
0.84 / 0.60 / 0.70 / 0.54 &
0.80 / 0.58 / 0.67 / 0.51 \\ 
\hline\noalign{\smallskip}
NCVR & 
0.80 / 0.81 / 0.80 / 0.67 &
0.81 / 0.75 / 0.77 / 0.64 &
0.80 / 0.72 / 0.76 / 0.61 \\ 
\hline\noalign{\smallskip}
European Census~~~~ & 
0.96 / 0.85 / 0.88 / 0.79 &
0.97 / 0.80 / 0.87 / 0.78 &
0.92 / 0.75 / 0.83 / 0.70 \\
\hline\noalign{\smallskip}
\end{tabular}
%\end{footnotesize}
\end{small}
\end{table*}

The re-identification results suggest that the approaches DP-BF and DP-DL-BF have low re-identification risk compared to BF and DL-BF approaches because 
%the frequencies of q-grams can be correctly identified as a lower number 
%of q-grams is mapped to a certain bit position. 
of differential privacy guarantees provided by DP-BF and DP-DL-BF approaches.
By flipping bits in BFs, 
the attack was not able to recognise the q-grams map to a certain bit position, 
thus avoiding the 1-to-1 or 1-to-many re-identifications. 
This is because many of the 
BFs contain different bit patterns leading each BF not being assigned 
to a plain text value.
Therefore, DL-DP-BF provides adequate privacy against frequency based privacy attacks, 
while achieving a high linkage quality as shown in Table~\ref{tab:linkage-quality}.

%Figure~\ref{fig:re-ident-prob-acm} shows the 
%the re-identification results for 
%the DBLP-ACM-C (top row) and European Census (bottom row) data sets. 
%As can be seen, when no differentially private (DP) noise is added ($p$ = 0) 
%in BFs in the BF and DL-BF approaches the attack identified
%frequent attribute values correctly for different attribute combinations.
%For the DBLP-ACM-C data set the attack correctly identified the 4 most frequent 
%author names correctly,

Further, we noted that as we increase the number of attributes the attack did not 
re-identify plain text values as not enough frequency information is available to
identify q-grams that are encoded in the BFs. Hence, 
conducting such cryptanalysis attacks upon BFs that are encoded with q-grams from
different attributes was unlikely to be successful. 
%We relegate the attack results with DBLP-ACM-C and European Census
%data sets to Fig.~\ref{fig:re-ident-prob-acm} in Appendix C.

%Further, as the results show, the attack did not 
%manage to correctly identify the attribute values encoded in BFs when
%bits are flipped according to DP noise.

\subsection{Ablation Study}

To evaluate the robustness of our approach, 
we ran experiments with different number of hash functions and different 
multi-party linkage scenarios including 2, 3, 5, 7 , and 10 DOs. 
In each multi-party linkage scenario, 
we randomly sample 50\% of tuples from each training data set for a DO. 

%\begin{figure*}[!t]
%  \centering
%    \includegraphics[width=0.33\textwidth]
%  {figures/all_dp_prob.eps}
%  \includegraphics[width=0.33\textwidth]
%  {figures/all_hash_funct.eps}
%  \includegraphics[width=0.33\textwidth]
%  {figures/all_parties.eps}
%  \caption{Average precision, recall, F-measure, and F*-measure results across all data sets.}
%  \label{fig:test2}
%\end{figure*}

Figure~\ref{fig:aba-runtime} shows runtime required for the training and 
classification phases of our approach with different flip probability ($p$),
number of hash functions ($k$), and number of DOs. As can be seen from this 
figure, the classification phase consumes more runtime compared to the training 
phase due to the additional runtime required by the LSH blocking technique.
Thus, we would like to highlight that the efficiency of the classification 
technique can be improved by using a more efficient blocking technique. 

The runtime did not change much with different $p$ values. 
This is because BFs are processed in the same way, 
according to Eq.~\ref{eqn:blip_rappor}, 
with different $p$ values in the noise addition step.
However, 
we noted the runtime increases with larger $k$ values as each q-gram in an attribute value needs to be hashed by large number of hash functions. 
We have seen the runtime required for the training phase increases linearly with an increasing number of DOs while runtime for classification by the LU increases exponentially. 
This is because in the training phase each DO performs training of their local models independently while the LU performs an exponential number of BF pair comparisons. 
For example, 
assuming each data set contains $n$ records and $b$ equal-sized blocks, 
a linkage of $d$ databases requires $(n/b)^d$ comparisons to be performed in the classification phase. 

Tables~\ref{tab:linkage-quality-p} and \ref{tab:linkage-quality-k} show linkage quality
results of our approach with various $p$ and $k$ values. As can be seen from
Table~\ref{tab:linkage-quality-p}, the linkage quality decreases slightly (5\% -- 12\% in
terms of F-measure) as $p$ grows. This is because a large value of $p$ can flip a high 
number of bits in BFs as we show in Table~\ref{tab:harden-blip}. As shown in
Table~\ref{tab:linkage-quality-k}, we noted that our approach resulted in
similar precision and recall with different $k$. This is because of the hashing of 
q-grams does not affect the training of the classification model in the training phase. 

\begin{table*}[t!]
    \addtolength{\tabcolsep}{-4.0pt}
\centering
  \caption{Average linkage quality results (P: Precision, R: Recall,
  F: F-measure, and F*: F*-measure) with different $k$ hash
  functions on different data sets. 
  \label{tab:linkage-quality-k}}
\begin{small}
\begin{tabular}{llll}
\hline
\multirow{2}{*}{Data set} & 
\multicolumn{3}{c}{Number of hash functions ($k$)}
\\ \cline{2-4}
 & \multicolumn{1}{c}{\begin{tabular}[c]{@{}c@{}}10\\ P / R / F / F*\end{tabular}} & \multicolumn{1}{c}{\begin{tabular}[c]{@{}c@{}}20\\ P / R / F / F*\end{tabular}} & \multicolumn{1}{c}{\begin{tabular}[c]{@{}c@{}}30\\ P / R / F / F*\end{tabular}} \\ 
 \hline\noalign{\smallskip}
DBLP-ACM-C &
0.94 / 0.78 / 0.85 / 0.74 ~~~~&
0.95 / 0.74 / 0.83 / 0.71 ~~~~& 
0.96 / 0.75 / 0.84 / 0.72 \\ 
\hline\noalign{\smallskip}
DBLP-ACM-D & 
0.95 / 0.75 / 0.83 / 0.72 &
0.93 / 0.75 / 0.82 / 0.70 & 
0.94 / 0.70 / 0.80 / 0.67 \\
\hline\noalign{\smallskip}
DBLP-Scholar-C &
0.99 / 0.55 / 0.71 / 0.55 &
0.98 / 0.52 / 0.68 / 0.51 & 
0.97 / 0.52 / 0.67 / 0.51 \\ 
\hline\noalign{\smallskip}
DBLP-Scholar-D & 
0.98 / 0.43 / 0.59 / 0.43 &
0.99 / 0.42 / 0.58 / 0.41 & 
0.98 / 0.40 / 0.56 / 0.39 \\ 
\hline\noalign{\smallskip}
iTune-Amazon & 
0.82 / 0.51 / 0.62 / 0.46 &
0.83 / 0.50 / 0.62 / 0.43 & 
0.81 / 0.48 / 0.61 / 0.44 \\
\hline\noalign{\smallskip}
Music-Brainz & 
0.99 / 0.73 / 0.84 / 0.72 &
0.98 / 0.72 / 0.83 / 0.70 & 
0.98 / 0.70 / 0.81 / 0.69 \\
\hline\noalign{\smallskip}
Amazon-Google &
0.85 / 0.63 / 0.72 / 0.57 &
0.84 / 0.60 / 0.70 / 0.54 & 
0.83 / 0.60 / 0.69 / 0.53 \\ 
\hline\noalign{\smallskip}
NCVR & 
0.85 / 0.83 / 0.81 / 0.69 &
0.79 / 0.80 / 0.80 / 0.68 & 
0.84 / 0.79 / 0.80 / 0.68 \\ 
\hline\noalign{\smallskip}
European Census~~~~ & 
0.98 / 0.83 / 0.89 / 0.80 &
0.97 / 0.81 / 0.88 / 0.79 & 
0.99 / 0.79 / 0.87 / 0.77 \\
\hline\noalign{\smallskip}
\end{tabular}
\end{small}
\end{table*}

Table~\ref{tab:linkage-quality-mp} displays the linkage quality results for different
numbers of DOs involved in the linkage process. As can be seen from the table, 
the linkage quality decreases (2\% -- 15\% in precision and 15\% -- 30\% in recall) 
when more databases are to be linked. This is due to the randomisation in DP noise
perturbation in BFs as different bits can be flipped in matching BFs in different 
databases. This suggests that a larger $p$ value can result in the overall linkage
quality decreasing, while the privacy of BFs is increased. Thus, a multi-party linkage 
project requires a careful optimisation of the $p$ value to achieve high linkage 
quality and privacy. 

\begin{table*}[t!]
    \addtolength{\tabcolsep}{-3.5pt}
\centering
  \caption{Average linkage quality results (P: Precision, R: Recall, F: F-measure,
    and F*: F*-measure) for different data sets with different number of database owners (DOs). 
  \label{tab:linkage-quality-mp}}
%\begin{small}
\begin{footnotesize}
\begin{tabular}{lccccc}
\hline\noalign{\smallskip}
Data set & 
\multicolumn{1}{c}{\begin{tabular}[c]{@{}c@{}}2\\ P / R / F / F*\end{tabular}} &
\multicolumn{1}{c}{\begin{tabular}[c]{@{}c@{}}3\\ P / R / F / F*\end{tabular}} &
\multicolumn{1}{c}{\begin{tabular}[c]{@{}c@{}}5\\ P / R / F / F*\end{tabular}} &
\multicolumn{1}{c}{\begin{tabular}[c]{@{}c@{}}7\\ P / R / F / F*\end{tabular}} &
\multicolumn{1}{c}{\begin{tabular}[c]{@{}c@{}}10\\ P / R / F / F*\end{tabular}} \\ 
\hline\noalign{\smallskip}
\begin{tabular}[l]{@{}l@{}}DBLP-\\ ACM-C\end{tabular} & 
0.95 / 0.78 / 0.85 / 0.74 ~~~& 
0.92 / 0.75 / 0.83 / 0.71 ~~~& 
0.85 / 0.72 / 0.77 / 0.63 ~~~& 
0.80 / 0.72 / 0.75 / 0.61 ~~~&
0.78 / 0.65 / 0.70 / 0.54 \\ 
\hline\noalign{\smallskip}

\begin{tabular}[l]{@{}l@{}}DBLP-\\ ACM-D\end{tabular} & 
0.96 / 0.75 / 0.84 / 0.72 & 
0.94 / 0.75 / 0.83 / 0.71 & 
0.91 / 0.70 / 0.79 / 0.65 & 
0.96 / 0.65 / 0.77 / 0.63 &
0.86 / 0.60 / 0.70 / 0.54 \\ 
\hline\noalign{\smallskip}

\begin{tabular}[l]{@{}l@{}}DBLP-\\ Scholar-C\end{tabular} &
0.99 / 0.56 / 0.71 / 0.55 & 
0.98 / 0.56 / 0.71 / 0.55 & 
0.98 / 0.50 / 0.66 / 0.49 & 
0.96 / 0.43 / 0.59 / 0.42 &
0.96 / 0.40 / 0.56 / 0.39 \\ 
\hline\noalign{\smallskip}

\begin{tabular}[l]{@{}l@{}}DBLP-\\ Scholar-D\end{tabular} &
0.98 / 0.50 / 0.66 / 0.49 & 
0.98 / 0.50 / 0.66 / 0.49 &
0.96 / 0.46 / 0.62 / 0.42 & 
0.95 / 0.40 / 0.56 / 0.39 &
0.90 / 0.38 / 0.53 / 0.36 \\ 
\hline\noalign{\smallskip}

\begin{tabular}[l]{@{}l@{}}iTune-\\ Amazon\end{tabular} & 
0.81 / 0.51 / 0.62 / 0.45 & 
0.80 / 0.45 / 0.57 / 0.40 & 
0.80 / 0.45 / 0.57 / 0.40 & 
0.84 / 0.38 / 0.52 / 0.35 &
0.83 / 0.35 / 0.49 / 0.32 \\ 
\hline\noalign{\smallskip}

\begin{tabular}[l]{@{}l@{}}Music-\\ Brainz\end{tabular} & 
0.99 / 0.73 / 0.84 / 0.72 & 
0.98 / 0.68 / 0.80 / 0.67 & 
0.99 / 0.65 / 0.78 / 0.64 & 
0.98 / 0.60 / 0.74 / 0.59 &
0.98 / 0.60 / 0.74 / 0.59 \\ 
\hline\noalign{\smallskip}

\begin{tabular}[l]{@{}l@{}}Amazon-\\ Google\end{tabular} & 
0.85 / 0.63 / 0.72 / 0.56 & 
0.85 / 0.58 / 0.68 / 0.52 & 
0.82 / 0.55 / 0.65 / 0.49 & 
0.81 / 0.38 / 0.51 / 0.34 &
0.81 / 0.38 / 0.51 / 0.34 \\ 
\hline\noalign{\medskip}

NCVR & 
0.80 / 0.82 / 0.81 / 0.68 & 
0.80 / 0.82 / 0.81 / 0.68 & 
0.82 / 0.79 / 0.80 / 0.67 & 
0.75 / 0.72 / 0.73 / 0.58 &
0.75 / 0.72 / 0.73 / 0.58 \\
\noalign{\medskip}
\hline\noalign{\smallskip}

\begin{tabular}[l]{@{}l@{}}European\\ Census\end{tabular} & 
0.97 / 0.82 / 0.88 / 0.79 & 
0.99 / 0.80 / 0.88 / 0.79 & 
0.97 / 0.71 / 0.81 / 0.69 & 
0.98 / 0.68 / 0.80 / 0.67 &
0.95 / 0.58 / 0.72 / 0.57 \\ 
\hline\noalign{\smallskip}
\end{tabular}
\end{footnotesize}
%\end{small}
\end{table*}

\section{Conclusion}
\label{sec:conclusion}
Deep learning has been used in record linkage due to its superior performance
in linking records. However, existing deep learning techniques cannot be used 
in privacy-preserving record linkage (PPRL) due to data availability issues.
For the first time, we proposed a deep learning-based privacy-preserving 
record linkage (PPRL) protocol that can be used to link sensitive databases 
held by different organisations with high linkage accuracy. The organisations
collaboratively train a deep learning model on deferentially private (DP) 
Bloom filters (BFs) which is then used by a linkage unit to classify unlabelled
BF pairs as matches and non-matches. Our experimental evaluation shows that the
proposed approach can achieve high linkage quality while providing strong 
privacy protection against privacy attacks. As future work, we aim to explore
how local DP can be used with an untrustworthy aggregator in the training 
phase of our approach. We also plan to investigate how transfer 
learning~\cite{kir22} can be used to address the problem of limited 
training data in PPRL.

% if have a single appendix:
%\appendix[Proof of the Zonklar Equations]
% or
%\appendix  % for no appendix heading
% do not use \section anymore after \appendix, only \section*
% is possibly needed

% use appendices with more than one appendix
% then use \section to start each appendix
% you must declare a \section before using any
% \subsection or using \label (\appendices by itself
% starts a section numbered zero.)
%

\iffalse
\appendices
\section{Proof of Theorem 5.1.}
%\subsection*{Appendix A}
% you can choose not to have a title for an appendix
% if you want by leaving the argument blank
\section{Description of F* Measure}
\section{Re-identification results on Other data sets}
% use section* for acknowledgment
\ifCLASSOPTIONcompsoc
  % The Computer Society usually uses the plural form
  \section*{Acknowledgments}
\else
  % regular IEEE prefers the singular form
  \section*{Acknowledgment}
\fi

The authors would like to thank...

\fi
% Can use something like this to put references on a page
% by themselves when using endfloat and the captionsoff option.
\ifCLASSOPTIONcaptionsoff
  \newpage
\fi

% trigger a \newpage just before the given reference
% number - used to balance the columns on the last page
% adjust value as needed - may need to be readjusted if
% the document is modified later
%\IEEEtriggeratref{8}
% The "triggered" command can be changed if desired:
%\IEEEtriggercmd{\enlargethispage{-5in}}

% references section

% can use a bibliography generated by BibTeX as a .bbl file
% BibTeX documentation can be easily obtained at:
% http://mirror.ctan.org/biblio/bibtex/contrib/doc/
% The IEEEtran BibTeX style support page is at:
% http://www.michaelshell.org/tex/ieeetran/bibtex/
\bibliographystyle{IEEEtran}
% argument is your BibTeX string definitions and bibliography database(s)
\bibliography{IEEEabrv,citations}

\vfill

% Can be used to pull up biographies so that the bottom of the last one
% is flush with the other column.
%\enlargethispage{-5in}

% that's all folks
\end{document}